\documentclass{chjaa}                  

\usepackage{graphicx,times}             
\input{epsf.sty}                        

\begin{document}

   \title{{\it Review:} The long-term survival chances of young massive star clusters}

   \volnopage{Vol.0 (2007) No.0, 000--000}      
   \setcounter{page}{1}           

   \author{Richard de Grijs
      \inst{1,2}\mailto{R.deGrijs@sheffield.ac.uk}
   \and Genevi\`eve Parmentier
      \inst{3}
      }

   \institute{Department of Physics \& Astronomy, The University of Sheffield,
   Hicks Building, Hounsfield Road, Sheffield S3 7RH, U. K.\\
   \and
             National Astronomical Observatories, Chinese Academy of Sciences,
             20A Datun Road, Chaoyang District, Beijing 100012, P. R. China\\
             \email{R.deGrijs@sheffield.ac.uk}
   \and
             Institute of Astrophysics \& Geophysics, Universit\'e de Li\`ege,
    All\'ee du 6 Ao\^{u}t, 17, Sart-Tilman (B5c), B-4000 Li\`ege,
    Belgium\\
          }

   \date{Received~~2007 month day; accepted~~2007~~month day}

   \abstract{We review the long-term survival chances of young massive
   star clusters (YMCs), hallmarks of intense starburst episodes often
   associated with violent galaxy interactions. We address the key
   question as to whether at least some of these YMCs can be
   considered proto-globular clusters (GCs), in which case these would
   be expected to evolve into counterparts of the ubiquitous old GCs
   believed to be among the oldest galactic building blocks. In the
   absence of significant external perturbations, the key factor
   determining a cluster's long-term survival chances is the shape of
   its stellar initial mass function (IMF). It is, however, not
   straightforward to assess the IMF shape in unresolved extragalactic
   YMCs. We discuss in detail the promise of using high-resolution
   spectroscopy to make progress towards this goal, as well as the
   numerous pitfalls associated with this approach. We also discuss
   the latest progress in worldwide efforts to better understand the
   evolution of entire cluster systems, the disruption processes they
   are affected by, and whether we can use recently gained insights to
   determine the nature of at least some of the YMCs observed in
   extragalactic starbursts as proto-GCs. We conclude that there is an
   increasing body of evidence that GC formation appears to be
   continuing until today; their long-term evolution crucially depends
   on their environmental conditions, however.   }

   \authorrunning{Richard de Grijs \& Genevi\`eve Parmentier}            
   \titlerunning{Massive star cluster formation and long-term survival}  

   \maketitle

\section{Introduction}           
\label{sect:intro}

Young, massive star clusters (YMCs) are the hallmarks of violent
star-forming episodes triggered by galaxy collisions and close
encounters. Their contribution to the total luminosity induced by such
extreme conditions completely dominates the overall energy output due
to the gravitationally-induced star formation (e.g., Holtzman et
al. 1992; Whitmore et al. 1993; O'Connell, Gallagher \& Hunter 1994;
Conti, Leitherer \& Vacca 1996; Watson et al. 1996; Carlson et
al. 1998; de Grijs et al. 2001, 2003a,b,c,d,e).

The question remains, however, whether or not at least a fraction of
the compact YMCs seen in abundance in extragalactic starbursts, are
potentially the progenitors of ($\ga 10$ Gyr) old globular cluster
(GC)-type objects -- although of higher metallicity than the
present-day GCs. If we could settle this issue convincingly, one way
or the other, such a result would have far-reaching implications for a
wide range of astrophysical questions, including our understanding of
the process of galaxy formation and assembly, and the process and
conditions required for star (cluster) formation. Because of the lack
of a statistically significant sample of YMCs in the Local Group,
however, we need to resort to either statistical arguments or to the
painstaking approach of case-by-case studies of individual objects in
more distant galaxies.

A variety of methods have been employed to address the issue of the
long-term survival of massive star clusters. The most promising and
most popular approach aimed at establishing whether a significant
fraction of an entire {\it population} of YMCs (as opposed to
individual objects) might survive for any significant length of time
(say, in excess of a few $\times 10^9$ yr) uses the ``cluster
luminosity function'', or its equivalent mass function (CLF, CMF), as
a diagnostic tool. In essence, the long-term survival of dense YMCs
depends sensitively on the low-mass section (below a few $M_\odot$) of
their stellar initial mass function (IMF). Clearly, assessing the
shape of the stellar IMF in unresolved extragalactic star clusters is
difficult, potentially ambiguous and riddled with pitfalls.
Nevertheless, and despite these difficulties, an ever increasing body
of observational evidence lends support to the scenario that GCs,
which were once thought to be the oldest building blocks of galaxies,
are still forming today.

In this review, we discuss the chances for YMCs to survive a Hubble
time of evolution and disruptive internal and external effects, and
thus to become equivalent objects to the old GCs seen in abundance in
the large spiral and elliptical galaxies in the local Universe. This
is a very topical area of research in the star cluster community.
Here, we will review the variety of view points expressed in the
community and address both the well-established cluster evolution
scenarios and the more controversial issues that regularly appear in
the literature. We start, in Sect. \ref{infant.sec}, by focusing on
the process of ``cluster infant mortality'', which in essence causes
the almost instantaneous disruption of some 90 per cent of a young
cluster population in the first $\sim 30$ Myr of their lifetime. We
then discuss the effects and observational signatures of star cluster
disruption on longer time-scales (Sect. \ref{disruption.sec}). In
Sect. \ref{individual.sec} we address the survival chances of
individual massive star clusters, for which we have access to the most
up-to-date and detailed observations. Whereas we discuss the internal
dynamics of individual star clusters to some extent, for more details
we refer the interested reader to the excellent review on this topic
by Meylan \& Heggie (1997). We continue, in Sect. \ref{clfs.sec}, by
scrutinising the evolution of entire cluster populations, in essence
by looking at their CLFs and CMFs. Here, we focus predominantly on the
evolution of {\it young} cluster systems; for an in-depth overview of
the properties, evolution and context of old GC systems we refer to
the recent review by Brodie \& Strader (2006). In Sect.
\ref{summary.sec} we provide a summary of our main conclusions and an
outlook to future developments in this very vibrant field.

\section{Infant Mortality}
\label{infant.sec}

Observations of increasing numbers of interacting and starburst
galaxies show a significantly larger number of young ($\la 10-30$ Myr)
star clusters than expected from a simple extrapolation of the cluster
numbers at older ages, under the assumption of a roughly constant star
cluster formation rate over the host galaxy's history, and taking into
account the observational completeness limits as well as the effects
of sample binning. Notable examples of galaxies in which this effect
is clearly seen include the Antennae interacting system, NGC 4038/9
(Whitmore 2004; Fall, Chandar \& Whitmore 2005; Mengel et al. 2005;
see also Whitmore, Chandar \& Fall 2006 for a presentation of earlier
results), M51 (Bastian et al. 2005), and NGC 6745 (de Grijs et
al. 2003c). This significant overdensity remains, even in view of the
presence of a recent burst of star cluster formation in many of these
galaxies. In addition, there has been a recent surge of interest in
the star cluster populations in the Magellanic Clouds (Rafelski \&
Zaritsky 2005; Chandar, Fall \& Whitmore 2006; Gieles, Lamers \&
Portegies Zwart 2007; de Grijs et al. 2007), which we can probe in
much more detail and to much fainter flux limits than the
extragalactic starburst cluster populations.

\subsection{Cluster Disruption on Short Time-Scales}

The current consensus is that there are most likely two types of
cluster disruption scenarios, which can account for these observations
on their own specific time-scales. These include an initial fast ($\la
10-30$ Myr) disruption mechanism that may be mass-independent -- at
least for masses in excess of $\sim 10^4 M_\odot$ (e.g., Bastian et
al. 2005; Fall et al. 2005; Fall 2006) -- and a subsequent secondary
(secular) disruption mechanism.

The proposed fast disruption mechanism, which is thought to
effectively remove up to 50 (Wielen 1971, 1988; Rafelski \& Zaritsky
2005; Goodwin \& Bastian 2006), 70 (Bastian et al. 2005; Mengel et
al. 2005; de Grijs et al. 2007) or even 90 per cent (Lada \& Lada
1991; Whitmore 2004; Whitmore et al. 2006; see also Lamers \& Gieles
2007) of the youngest, short-lived clusters from a given cluster
population, has been coined cluster ``infant mortality'' (Whitmore
2004). The observational effects and their consequences had, however,
already been discussed in detail much earlier, both in Galactic (e.g.,
Wielen 1971, 1988; Lada \& Lada 1991, 2003) and in extragalactic
environments (Tremonti et al. 2001).

In particular, based on ultraviolet {\sl Hubble Space Telescope
(HST)}/STIS spectroscopy of the YMCs and diffuse background light in
the dwarf starburst galaxy NGC 5253, the latter study suggested that
star clusters may have been forming continuously in this galaxy, and
then dissolve on $\sim$ 10--20 Myr time-scales (based on arguments
related to tidal effects worked out in detail in Kim, Morris \& Lee
1999), dispersing their stars into the field star population. The main
observational evidence lending support to this suggestion comes from
the composite spectrum of eight YMCs versus that of the diffuse
``inter-cluster'' field star population covered by their long-slit
spectra. While the former exhibits so-called P Cygni features in their
N{\sc v}, S{\sc iv} and C{\sc iv} lines, characteristic of stellar
winds generated by massive young O-type stars, the latter spectrum
lacks these stellar wind features.

Tremonti et al. (2001) conclude from the difference between the
composite cluster and field-star spectra that the NGC 5253 field star
population contains stars at a more advanced evolutionary stage than
the clusters they sampled; they suggest that a straightforward
interpretation of this result is that stars tend to form in star
clusters, while a fraction of them populate the field after the
clusters have dissolved on 10--20 Myr time-scales. They point out that
a similar scenario seems feasible for NGC 1569 as well, based on the
{\sl HST}/WFPC2 photometry of Greggio et al. (1998). These latter
authors find that the stellar population of NGC 1569 is composed of
recently formed YMCs and resolved field stars with ages in excess of
10 Myr [see also Harris et al. (2001) for a similar conclusion based
on the resolved diffuse stellar population in M83]. In a follow-up
paper, Chandar et al. (2005) compare the {\sl HST}/STIS ultraviolet
spectral signatures of both the star clusters and the diffuse
intercluster light originating in the galactic disks of 12 local
starburst galaxies. They show that in 11 of their 12 sample galaxies
the ``field'' spectra lack the O-star signatures observed in the
(young) star cluster spectra; the ``field'' spectra are, instead,
dominated by B-type stellar features (the only exception to this rule
is Henize 2-10, which exhibits the youngest field star population
among their sample galaxies, including O-star signatures). Thus,
Chandar et al. (2005) conclude that, under the assumption that the
field star population is composed of dissolved clusters, cluster
dissolution needs to happen on time-scales of 7--10 Myr. If, instead,
the field is composed of unresolved young clusters, these need to be
less massive than a few $\times 10^2 M_\odot$, in order to lack O
stars at the youngest ages.

Similarly, Pellerin et al. (2006) conclude, based on resolved
colour-magnitude diagrams obtained with {\sl HST}/Advanced Camera for
Surveys (ACS) that the background field stars in the nearby galaxy NGC
1313 are young, massive ($m_\ast \ga 7 M_\odot$) B-type stars. Since
such stars tend to be born in clusters and have lifetimes of 5 to 25
Myr, they conclude that this implies that significant cluster ``infant
mortality'' must be at work in this environment. They also suggest
that this is a plausible scenario to explain the presence of a B-type
stellar population exciting the diffuse ionised gas of normal galaxies
(Hoopes et al. 2001), and for the bright diffuse UV emission observed
to account for some 80 per cent of the UV emission in starburst
galaxies (Meurer et al. 1995).

The rationale for the cluster ``infant mortality'' interpretation for
the Antennae galaxies, rather than a recent large-scale, intense burst
of star cluster formation (which might mimic the observations as
well), was provided in detail by Whitmore (2004), and repeated by Fall
et al. (2005) and Whitmore et al. (2006). It consists of two
interlinked arguments: (i) the Antennae galaxies have been interacting
for much longer than the median cluster age (the time-scales involved
are a few $\times 10^8$ versus $\sim 10^7$ yr, respectively), and (ii)
the cluster age distribution is similar across different areas of the
galaxies, irrespective of the intensity of the interaction, and on
spatial scales much larger than those traveled by the sound speed on
time-scales on the order of the median cluster age.

\subsection{Disruption Mechanisms}

Star clusters are subject to a variety of internal and external
mechanisms that, under the appropriate conditions, will
gravitationally unbind and subsequently disrupt them. These effects
include (see Mengel et al. 2005), approximately as a function of
increasing time-scale, (i) formation in a marginally bound state (see
also the review by Mac Low \& Klessen 2004), (ii) rapid removal of the
intracluster gas due to adiabatic or explosive expansion driven by
stellar winds or supernova activity, typically on time-scales much
shorter than the proto-cluster dynamical crossing time, (iii) mass
loss due to normal stellar evolution (including the effects of stellar
winds and supernova explosions), (iv) internal two-body relaxation
effects, leading to dynamical mass segregation and the preferential
ejection of lower-mass stars, hence altering the cluster's stellar
mass function, (v) release of energy stored in a significant fraction
of primordial ``hard'' binary systems, and (vi) tidal and
gravitational effects due to interactions with other significant mass
components, spiral arms, bulge or disc shocking, and dynamical
friction.

The general consensus emerging from recent studies into these effects
is that rapid gas removal from young star clusters, which could leave
them severely out of virial equilibrium, would be conducive to
subsequent cluster disruption (Vesperini \& Zepf 2003; Bastian et
al. 2005; Fall et al. 2005). The efficiency of this process will be
enhanced if a cluster's star-formation efficiency (SFE) is less than
about 30 per cent, independent of the mass of the cluster (see
e.g. Lada, Margulis \& Dearborn 1984; Goodwin 1997a,b; Adams 2000;
Geyer \& Burkert 2001; Kroupa \& Boily 2002; Boily \& Kroupa 2003a,b;
Fellhauer \& Kroupa 2005; Bastian \& Goodwin 2006; Parmentier \&
Gilmore 2007, their fig. 1). Goodwin \& Bastian (2006) show that this
type of cluster destruction occurs in 10--30 Myr (see also Kroupa \&
Boily 2002; Lada \& Lada 2003; and the discussion in Lamers \& Gieles
2007). The consequence of this is that clusters will expand rapidly,
in order to attain a new virial equilibrium, and hence disappear below
the observational detection limit on a similar time-scale (see also
Bastian et al. 2005; Fall et al. 2005). This might be the reason for
Mengel et al.'s (2005) conclusion that the average cluster size in the
Antennae system decreases with age, for ages up to $\sim 10$
Myr. However, they also note that the most extended clusters might be
more sensitive to dissolution due to tidal effects, and to mass loss
as a result of stellar winds (see also Vesperini \& Zepf 2003), and
hence the observed sample may maintain an apparently small average
size. Gieles et al. (2005) also noticed that, among their M51 cluster
sample, there are no old clusters with large radii. Although this
would suggest that large clusters are more easily disrupted over time,
and thus that tidal shocks may be dominating cluster disruption, their
result is hardly conclusive, given the observational uncertainties. On
the other hand, in the Large Magellanic Cloud (LMC) it is well known
that both the upper envelope to the size distribution of the star
cluster population and its spread increases with increasing age (e.g.,
Elson, Freeman \& Lauer 1989; Elson 1991, 1992; de Grijs et al. 2002c;
Mackey \& Gilmore 2003). The difference in this sense between the LMC
and the Antennae star cluster system might lie, apart from the obvious
difference in the observational detection limit, in the significantly
higher-density disk of the Antennae system compared to that of the
LMC, so that one might expect external effects due to disk shocking to
be more important in the Antennae galaxies.

The above scenario is, in essence, the key argument proposed by Hills
(1980) to explain the expansion of the Galactic OB associations, where
the energy and momentum output of massive stars causes an association
to expand, and ultimately dissolve. However, the problem is likely
more complex than outlined here. Boily \& Kroupa (2003a) suggest that
in order to form bound clusters with a low average SFE, the SFE must
peak in the cluster core, thus naturally leading to a concentration of
stars (see also Tenorio-Tagle et al. 1986; Adams 2000).

For completeness, we should point out that many of the youngest star
clusters observed in environments like those of the Antennae galaxies
might simply be unbound associations. Mac Low \& Klessen (2004, and
references therein) reviewed a star-formation scenario in which the
driving force is provided by supersonic turbulent convergence flows in
the interstellar medium, which might lead to high-intensity star
formation in {\it unbound} associations (see also Mengel et al. 2005).

\subsection{Cluster mass (in)dependence?}

\begin{figure}
   \plottwo{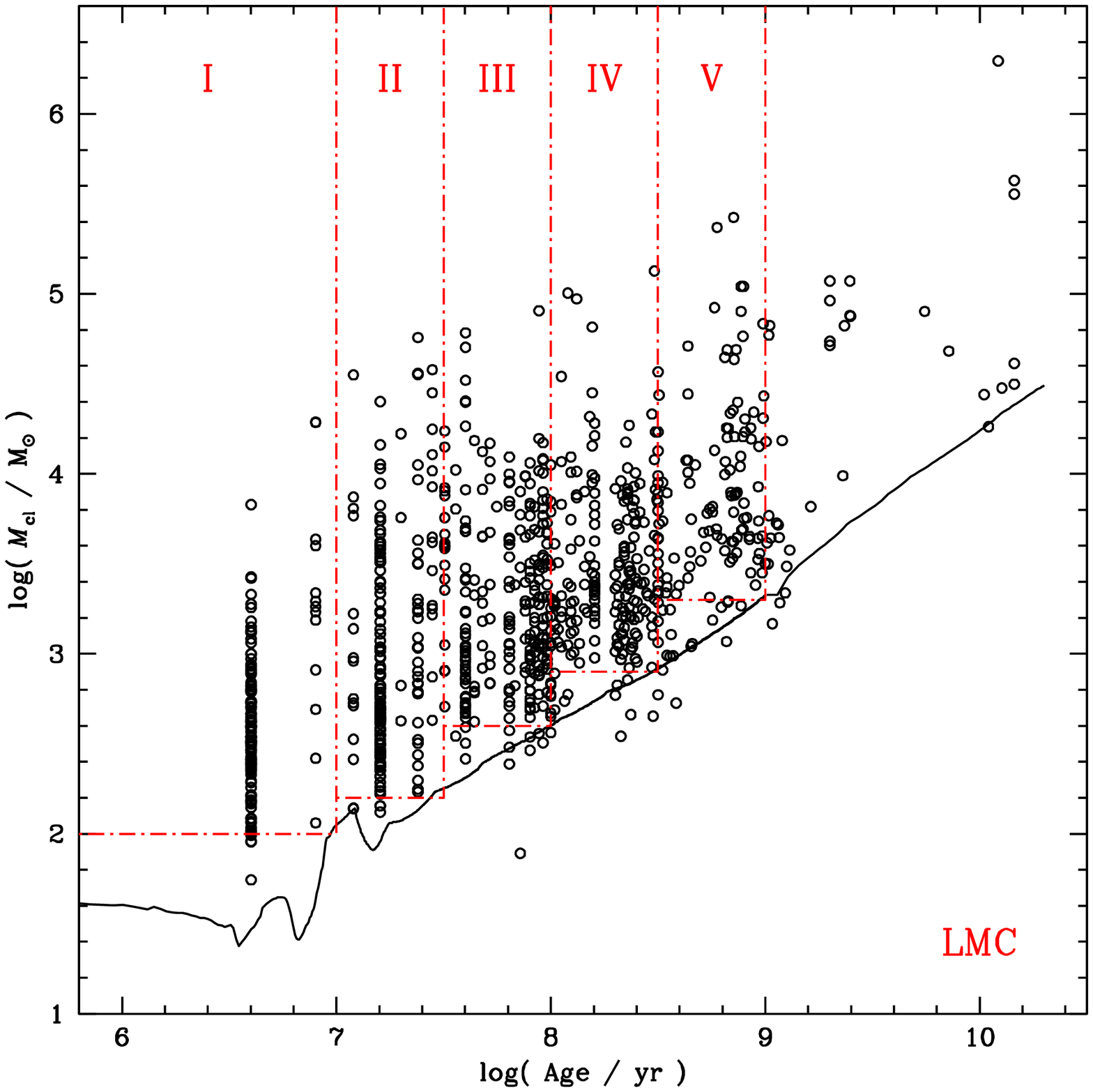} {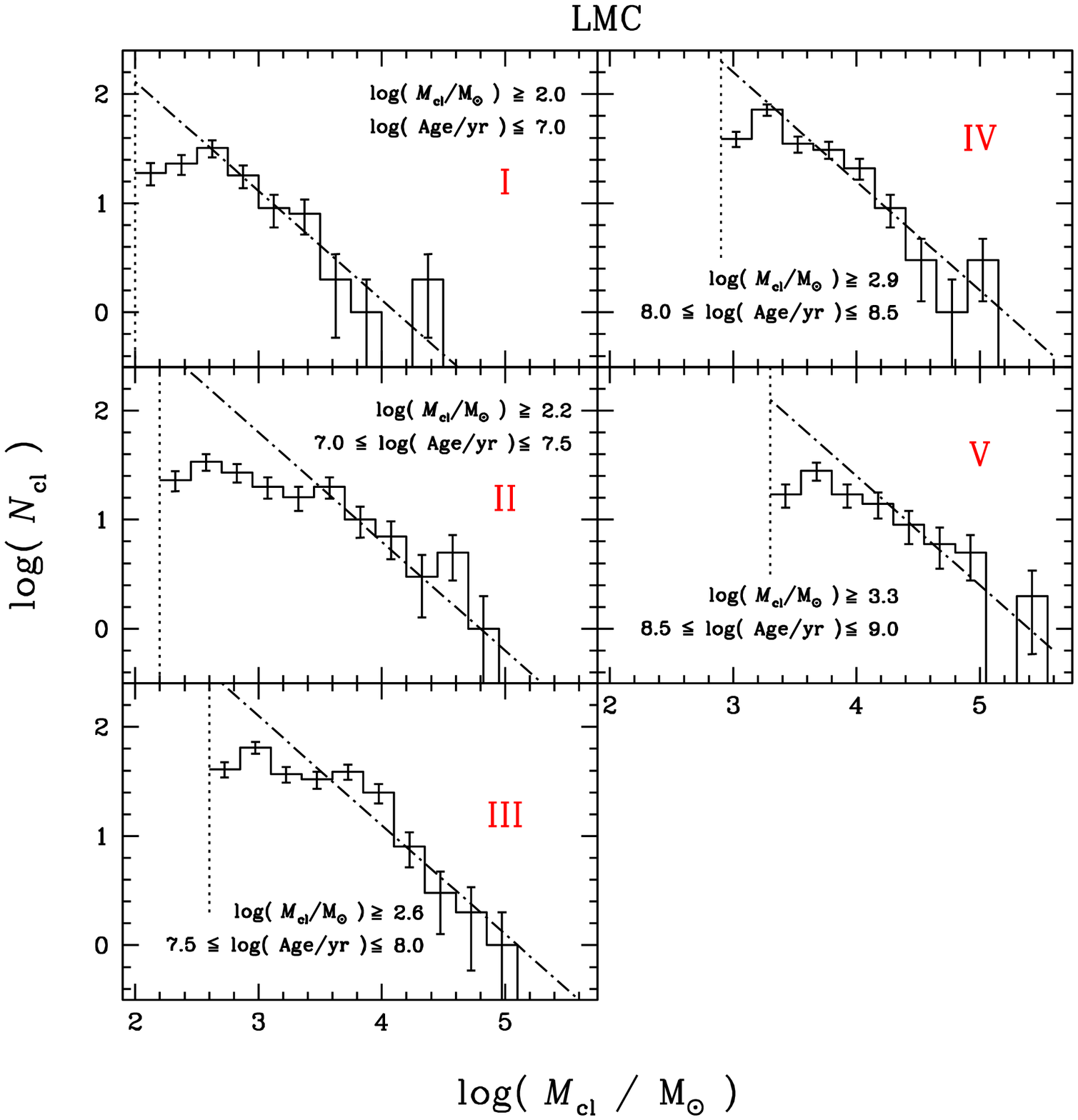}
   \caption{{\it Left:} Distribution of the LMC cluster sample of de
Grijs et al. (2007) in the (age versus mass) plane. Overplotted is the
expected detection limit based on stellar population synthesis for a
50 per cent completeness limit of $M_V = -3.5$ mag, assuming no
extinction. For a nominal extinction of $A_V = 0.1$ mag, the detection
limit will shift to higher masses by $\Delta \log( m_{\rm cl}
[M_\odot] ) = 0.04$. The features around 10 Myr are caused by the
appearance of red supergiants in the models. The age limits used to
generate the different panels in the right-hand panels are shown as
the vertical dash-dotted lines; the various subsets are also
cross-linked between the figures using roman numerals. The horizontal
dash-dotted lines indicate the 50 per cent completeness limits in mass
for each of the age-selected subsamples. {\it Right:} CMFs for
statistically complete LMC cluster subsamples. Age and mass ranges are
indicated in the panel legends; the vertical dotted lines indicate the
lower mass (50 per cent completeness) limits adopted. Error bars
represent simple Poissonian errors, while the dash-dotted lines
represent CMFs with spectral index $\alpha = 2$, shifted vertically to
match the observational data.}
   \label{lmc.fig}
\end{figure}

Bastian et al. (2005) attempted a first observational assessment of
the dependence of the cluster infant mortality on cluster mass in
M51. Their results seem to indicate that the effect of infant
mortality is largely independent of mass, although close inspection of
their Fig. 14 shows that this may not be as strongly supported for the
lower masses ($m_{\rm cl} \la 10^4 M_\odot$) as for the higher-mass
clusters (see also de Grijs et al. 2007). However, in view of the
large uncertainties, we cannot easily quantify the importance of a
possible mass dependence in M51. Fall et al. (2005; see also Fall
2006), in a more carefully worded statement, conclude that the shape
of the age distribution of the YMCs in the Antennae galaxies is nearly
independent of mass, at least for $m_{\rm cl} > 3 \times 10^4
M_\odot$. This is equivalent to the statement that the effects of
cluster infant mortality are largely mass independent above this mass
limit.

Such a result would be expected if, on average, the SFE of YMCs is
mass-independent. Goodwin \& Bastian (2006; also S. P. Goodwin, priv.
comm.) determined approximate SFEs for their sample of high-mass
clusters, and compared these to photometric mass estimates. They
mention in passing that the SFEs they determine for their sample
clusters are to first order mass-independent. The physical importance
of such a result, if this can be confirmed more robustly, implies that
the turbulent structure of the molecular cloud YMC progenitors is
scale-independent (e.g., Elmegreen 2002) and thus that the higher-mass
clusters are simply scaled-up versions of their lower-mass
counterparts.

We recently re-analysed the LMC cluster population in detail (de Grijs
et al. 2007; see Fig. \ref{lmc.fig}), thereby focusing on the effects
of infant mortality, and concluded that there appears to be a
mass-dependent infant mortality rate, at least for masses below a few
$\times 10^3 M_\odot$. In view of the arguments presented above in
favour of a mass-independent YMC SFE, this observation seems to imply
that clusters below this mass limit might not simply be scaled-down
versions of the higher-mass YMCs, but that the effective cluster SFE
does in fact depend on mass, thus establishing a clearer distinction
between gravitationally bound and unbound objects at the low-mass
extreme (see also Weidner et al. 2007 for a theoretical perspective).

In the right-hand panels of Fig. \ref{lmc.fig} we show the CMFs for
statistically complete subsamples of LMC clusters (taken from de Grijs
\& Anders 2006; de Grijs et al. 2007), as indicated in the left-hand
panel. We draw the reader's attention to the significant change in the
CMF slope between the youngest two age bins, which cannot be explained
satisfactorily by the difference in our mass sampling and the
completeness fractions only (de Grijs et al. 2007; based on extensive
Monte Carlo simulations of the observations). Instead, this is clear
observational evidence of mass-dependent cluster infant mortality from
clusters younger than 10 Myr in the first age bin to clusters in the
age range between 10--30 Myr in the second.

\section{Cluster disruption beyond the infant mortality phase}
\label{disruption.sec}

\subsection{General framework}

Those clusters that survive the infant mortality phase will be subject
to the processes driving longer-term star cluster dissolution. The
longer-term dynamical evolution of star clusters is determined by a
combination of internal and external time-scales. The free-fall and
two-body relaxation time-scales, which depend explicitly on the
initial cluster mass density (e.g., Spitzer 1958; Chernoff \& Weinberg
1990; de la Fuente Marcos 1997; Portegies Zwart et al. 2001), affect
the cluster-internal processes of star formation and mass
redistribution through energy equipartition, leading to mass
segregation and, eventually, core collapse. While the internal
relaxation process will, over time, eject both high-mass stars from
the core (e.g., due to interactions with hard binaries; see Brandl et
al. 2001; de Grijs et al. 2002a) and lose lower-mass stars from its
halo through diffusion (e.g., due to Roche-lobe overflow), the
external processes of tidal disruption, disk and bulge shocking, and
stripping by the surrounding galactic field (see, e.g., De Marchi,
Pulone, \& Paresce 2006 for a detailed recent observational study in
this area) are in general more important for the discussion of this
disruption phase of star clusters. Tidal disruption is enhanced by
stellar evolution, leading to mass loss by winds and/or supernova
explosions, which will further reduce the stellar density in a
cluster, and thus make it more sensitive to external tidal forces.

From the bimodal age distribution of (young) open and (old) globular
clusters in the Milky Way, Oort (1957) concluded that disruption of
Galactic star clusters must occur on time-scales of $\sim 5 \times
10^8$ yr (see also Wielen 1971, 1988; Battinelli \& Capuzzo-Dolcetta
1991; Lamers et al. 2005b; Lamers \& Gieles 2006, 2007; Piskunov et
al. 2006). Roughly simultaneously, Spitzer (1958) derived an
expression for the disruption time-scale as a function of a cluster's
mean density, $\rho_c$ ($M_\odot$ pc$^{-3}$): $t_{\rm dis} = 1.9
\times 10^8 \rho_c$ yr, for $2.2 < \rho_c < 22 \; M_\odot$
pc$^{-3}$. More advanced recent studies, based on {\it N}-body
modeling, have shown that the cluster disruption time-scale is
sensitive to the cluster mass, the fraction of binary (or multiple)
stars, the ambient density (and hence the cluster's galactocentric
distance), its orbital velocity, and the IMF adopted (e.g., Aguilar
1988; Chernoff \& Weinberg 1990; de la Fuente Marcos 1997; Meylan \&
Heggie 1997; Baumgardt \& Makino 2003; Lamers, Gieles \& Portegies
Zwart 2005a; Gieles et al. 2005). However, Gieles et al. (2005)
showed, for the M51 cluster system (see also Sect. \ref{m51.sec}),
that the mass dependence of the cluster disruption time-scale, within
a given galaxy, is the dominant effect, with the other effects being
of second-order importance; the ambient density appears to be a major
driver of the variation of the characteristic cluster disruption
time-scale among galaxies (Portegies Zwart, Hut \& Makino 1998;
Portegies Zwart et al. 2002; Baumgardt \& Makino 2003; Lamers et
al. 2005a).

Boutloukos \& Lamers (2003) derived an empirical relation between the
disruption time and a cluster's initial mass from observations of a
small but diverse sample of galaxies containing rich cluster systems:
the Milky Way (i.e., the solar neighbourhood), the Small Magellanic
Cloud (SMC), M33 and the inner spiral arms of M51. Similar analyses
have since been published for M82's fossil starburst region ``B'' (de
Grijs et al. 2003a, 2005b), NGC 3310 and NGC 6745 (de Grijs et
al. 2003c), NGC 5461, NGC 5462 and NGC 5471 (Chen, Chu \& Johnson
2005), and in more detail for the solar neighbourhood (Lamers et
al. 2005b; Lamers \& Gieles 2006; see also Piskunov et al. 2006), M51
(Gieles et al. 2005), the SMC (Chiosi et al. 2006), and the LMC (de
Grijs \& Anders 2006). Boutloukos \& Lamers (2003) showed, based on an
analysis of the mass and age distributions of magnitude-limited
samples of clusters, that the empirical disruption time of clusters
depends on their initial mass\footnote{It is well-established,
however, that the disruption time-scale does not only depend on mass,
but also on the initial cluster density and internal velocity
dispersion. Following Boutloukos \& Lamers (2003), however, we point
out that if clusters are approximately in pressure equilibrium with
their environment, we can expect the density of all clusters in a
limited volume of a galaxy to be roughly similar, so that their
disruption time-scale will predominantly depend on their (initial)
mass, with the exception of clusters on highly-eccentric orbits.},
$M_{\rm i}$, as
\begin{equation}
t_{\rm dis} = t_4^{\rm dis} \Bigl(\frac{M_{\rm i}}{10^4
M_\odot}\Bigr)^{\gamma} \quad ,
\label{tdis}
\end{equation}
where $t_4^{\rm dis}$ is the disruption time of a cluster of initial
mass $M_{\rm i}=10^4 M_\odot$. The value of $\gamma$ is approximately
the same in these four galaxies, $\gamma = 0.62 \pm 0.06$ (see also
Gieles et al. 2005; Lamers et al. 2005b; Rafelski \& Zaritsky 2005;
Elmegreen et al. 2006; Lamers \& Gieles 2006, 2007). However, the
characteristic disruption time-scale, $t_4^{\rm dis}$, varies widely
in the different galaxies. The disruption time-scale is
shortest\footnote{Note that this time-scale was based on the
assumption of an underlying mass function of the form $N(m_{\rm cl})
\propto m_{\rm cl}^{-\alpha}$, with $\alpha = 2$ (cf. Boutloukos \&
Lamers 2003). However, this assumption is unlikely to be correct for
the M82 B cluster system, thus casting doubt on the very short
characteristic cluster disruption time-scale -- the shortest known of
any galactic disk region -- thus derived (de Grijs et al. 2005b; see
also Sects. \ref{m51.sec} and \ref{clfs.sec}).} in M82 B, $\log(
t_4^{\rm dis} {\rm yr}^{-1} ) \simeq 7.5$ (de Grijs et al. 2003a); it
is longest in the LMC, $\log( t_4^{\rm dis} {\rm yr}^{-1} ) = 9.9 \pm
0.1$ (Boutloukos \& Lamers 2002; de Grijs \& Anders 2006), and SMC,
$\log( t_4^{\rm dis} {\rm yr}^{-1} ) = 9.9 \pm 0.2$ (de la Fuente
Marcos 1997; Boutloukos \& Lamers 2003). This is not unexpected,
considering the low-density environment of the Magellanic Clouds --
and their relative paucity of giant molecular clouds compared to the
Milky Way -- in which these clusters are found (van den Bergh \&
McClure 1980; Elson \& Fall 1985, 1988; Hodge 1987, 1988; Lamers et
al. 2005a; see also Krienke \& Hodge 2004, for the NGC 6822 cluster
system).

Even with recent improvements (e.g., Gieles et al. 2005) to the simple
model of Boutloukos \& Lamers (2003), the characteristic cluster
destruction time-scales resulting from a more sophisticated,
non-instantaneous destruction process are very similar to those based
on the simple method; see Lamers et al. (2005a) for a detailed
comparison, in particular their table 1.

\subsection{Environmental Impact}
\label{m51.sec}

While there seems to be a very strong mass dependence of the
characteristic star cluster disruption time-scale in a given galaxy,
when comparing cluster systems from different galaxies the ambient
density becomes an important secondary effect (Portegies Zwart et
al. 1998, 2002; Baumgardt \& Makino 2003; Lamers et al. 2005a).
Baumgardt \& Makino (2003) and Portegies Zwart et al. (1998, 2002),
based on {\it N}-body simulations in the Galactic halo tidal field
(represented as a logarithmic gravitational potential), and Lamers et
al. (2005a), based on an empirical approach, showed that the
characteristic disruption time-scale in a given galaxy depends on the
initial mass and the ambient density, $\rho_{\rm amb}$, as
\begin{equation}
t_4^{\rm dis} \; ({\rm yr}) \propto M_{\rm i}^\gamma \; (M_\odot) \;
\rho_{\rm amb}^{-0.5} \; (M_\odot {\rm pc}^{-3}) .
\label{tdis2}
\end{equation}
Lamers et al. (2005a; their fig. 4) confront the theoretical {\it
N}-body predictions with their empirically derived disruption
time-scales, and show that both approaches are in reasonable
agreement, at least for the star cluster systems in the solar
neighbourhood, the SMC and M33. However, the {\it N}-body simulations
overpredict the cluster disruption time-scale significantly both for
the clusters in M51 [$\log( t_4^{\rm dis} {\rm yr}^{-1} ) = 7.85 \pm
0.22$; Lamers et al. 2005a; see also Boutloukos \& Lamers 2003], and
also for the YMCs in M82 B (de Grijs et al. 2005b). In both cases, the
{\it N}-body simulations overpredict the empirically derived
characteristic cluster disruption time-scales by a factor of $\sim
10$--15.

Gieles et al. (2005) explored whether the empirical result could have
underestimated the actual time-scale because of the assumption adopted
by Boutloukos \& Lamers (2003) and Lamers et al. (2005a) of a constant
cluster formation rate, despite the fact that M51 is clearly
interacting with its smaller companion galaxy, NGC 5195. This would
naturally lead to periods of enhanced cluster formation around the
time of closest approach between both galaxies, and thus render the
assumption of a constant cluster formation rate rather questionable.
In addition, Gieles et al. (2005) also take into account the effects
of cluster infant mortality (see Sect. \ref{infant.sec}), in essence
by disregarding all clusters younger than $\sim 10^7$ yr in their
sample; if they were to include these YMCs, this would artificially
shorten the disruption time-scale. Due to the lack of a reliable
method to distinguish between bound and unbound clusters at these
young ages, this is a reasonable approach. The latter correction
appears of the greatest importance for the M51 system; their new
estimates for the characteristic disruption time-scale, based on the
clusters older than $10^7$ yr, is $t_4^{\rm dis} = 1.0^{+0.6}_{-0.5}
\times 10^8$ yr for a constant cluster formation rate, and $t_4^{\rm
dis} \simeq 2.0 \times 10^8$ yr for both a cluster formation rate that
has been increasing linearly for the past Gyr (assuming a few
different models for the shape of the cluster formation rate during
this period), and a cluster formation rate undergoing exponentially
decaying bursts around the times of closest approach between the two
galaxies. These values are still a factor of $\sim 5$ below the
theoretically predicted time-scales. We note that Lamers et
al. (2005b) found a similar discrepancy for the Galactic open clusters
in the solar neighbourhood. In a very recent contribution, Lamers \&
Gieles (2007) show that if one takes the combined effects of stellar
evolution, the underlying galactic tidal field, and perturbations by
spiral arms and giant molecular clouds properly into account, this
discrepancy can be understood. They show that by including the
significant impact of cluster disruption by giant molecular clouds
(shown to be an order of magnitude more important than that of the
spiral arms, and just as important as the other effects combined), the
theoretically predicted cluster disruption time-scales shorten
dramatically, and the discrepancy between the theoretically and
empirically derived time-scales disappears.

For M82 B (de Grijs et al. 2003a), we did not have to correct for the
effects of infant mortality because of the greater average age of the
cluster system (and the much smaller number of very young clusters in
this region). In addition, we treated the periods before, during and
after the burst of cluster formation (which occurred $\sim 10^9$ yr
ago) separately, by assuming constant cluster formation rates during
each of these epochs. We note that the most crucial assumption
underpinning our disruption time-scale estimate is that of the initial
cluster mass distribution. For the M82 B time-scale -- following
Lamers et al. (2005a) -- we used an initial power-law CMF, with
spectral index $\alpha = -2$. However, we noted in de Grijs et
al. (2003d; see also Sect. \ref{clfs.sec}) that the observed CMF of
the approximately coeval cluster population in M82 B resembles
Vesperini's (1998) (quasi-)equilibrium log-normal CMF relatively
closely, for cluster masses down to $\log( m_{\rm cl} [M_\odot] )
\simeq 4.4$ (i.e., at our 100 per cent observational completeness
limit). The implication of our assumption of ``instantaneous
disruption'' only constrains $t^{\rm dis}_4$ to $t^{\rm dis}_4 \la
10^9$ yr, i.e., less than the present age of the clusters formed
simultaneously in the burst of cluster formation. Therefore, if the
initial CMF of the {\it bound} clusters comprising the observed M82 B
CMF at its current age is more similar to a log-normal distribution
than to a power law (Sect.  \ref{clfs.sec}; see also de Grijs et
al. 2005b), we cannot constrain the characteristic disruption
time-scale to better than $t^{\rm dis}_4 \la 10^9$ yr.

Based on the M51 cluster age and mass determinations of Bastian et
al. (2005; N. Bastian \& R. Scheepmaker, priv. comm.) we constructed
statistically complete CMFs covering intervals of 0.5 dex in $\log(
{\rm Age}\; {\rm yr}^{-1} )$. The M51 cluster subsamples in the age
ranges $\log( {\rm Age \; yr}^{-1} ) \le 7.0$, $7.0 < \log( {\rm Age
\; yr}^{-1} ) \le 7.5$, and $7.5 < \log( {\rm Age \; yr}^{-1} ) \le
8.0$ are all consistent with power laws down to the 50 per cent
observational completeness limits, at $\log( m_{\rm cl} [M_\odot] )
\simeq 3.3, 3.8$ and 4.1, respectively. It is clear, therefore, that a
similar explanation as for M82 B for the short characteristic
disruption time-scale cannot feasibly be invoked for the M51 clusters.
Instead, Gieles et al. (2005) suggest that this may be due to (i)
variations, in particular a low-mass cut-off, in the stellar IMF, (ii)
significant external perturbations, (iii) YMC formation in non-virial
equilibrium conditions, or to (iv) variations in the clusters' central
concentration (see also Lamers et al. 2005b in the context of the
Galactic open clusters in the solar neighbourhood).

As we will discuss in detail in Sect. \ref{individual.sec}, it is
unlikely that an entire YMC population is affected by a low-mass
cut-off to its IMF, although this may speed up cluster dissolution in
individual cases. The {\it N}-body models of Portegies Zwart et
al. (1998, 2002) and Baumgardt \& Makino (2003) assume a smooth
underlying gravitational potential; the introduction of additional
external perturbations by, e.g., giant molecular clouds or spiral-arm
passages will shorten the typical disruption time-scale (e.g.,
Ostriker, Spitzer \& Chevalier 1972; Terlevich 1987; Theuns 1991;
Gieles et al. 2006a,b; Lamers \& Gieles 2006, 2007; Gieles 2007; see
also Lamers et al. 2005b). Similarly, after the initial fast infant
mortality phase, the surviving clusters will have lost the gas
comprising their interstellar medium, and will expand in an attempt to
regain virial equilibrium as a consequence (see Sect.
\ref{expulsion.sec} for a detailed discussion). It is at this stage
that they are most vulnerable to additional tidal disruption, which
will thus also speed up their dissolution (e.g., Bastian \& Goodwin
2006; Goodwin \& Bastian 2006; and references therein). The efficiency
of this process will be enhanced if the YMCs are initially more
extended than assumed in the {\it N}-body models, which generally
adopt cluster concentrations similar to those of the Galactic GCs.

While all of these effects may operate in any galactic environment,
the M51 clusters discussed by Gieles et al. (2005), as well as the
massive clusters in M82 B (de Grijs et al. 2005b; and references
therein) and the Galactic open clusters in the solar neighbourhood
(Lamers et al. 2005b), are located in the dense inner regions of their
host galaxies, where one would expect any additional perturbing
effects to be most efficient. It may therefore simply be the case that
the {\it disk} regions of M51, M82 B and the solar neighbourhood
provide unsuitable conditions for proto-GCs to survive for a Hubble
time, while YMCs formed in the lower-density halo regions stand a
better chance of long-term survival.

\section{The Long-term Evolution of Individual Star Clusters}
\label{individual.sec}

The crucial question remains, however, whether or not at least {\it
some} of the YMCs observed in extragalactic starbursts might survive
to become (possibly somewhat more metal-rich) counterparts of the
Galactic GCs when they reach a similar age. If we could resolve this
issue convincingly, one way or the other, the implications would be
far-reaching for a wide range of astrophysical questions. On galactic
scales, on the one hand, this would improve -- or confirm -- our
understanding of how galaxy formation, assembly and evolution
proceeds, and what the process and necessary conditions are for star
(cluster) formation. On the other hand, on the scales of the
individual cluster stars, and possibly as a function of the
environment in which the clusters formed, star cluster survival for a
Hubble time sets tight constraints on the slope and possible low-mass
cut-off of the stellar IMF, as we will discuss in more detail below.

The evolution to old age of young clusters depends crucially on their
stellar IMF. If the IMF slope is too shallow, i.e., if the clusters
are significantly deficient in low-mass stars compared to, e.g., the
solar neighbourhood, they will likely disperse within about a Gyr of
their formation (e.g., Chernoff \& Shapiro 1987; Chernoff \& Weinberg
1990; Goodwin 1997b; Smith \& Gallagher 2001; Mengel et al. 2002;
Kouwenhoven, de Grijs \& Goodwin, in preparation). As a case in point,
Goodwin (1997b) simulated the evolution of $\sim 10^4 - 10^5 M_\odot$
YMCs similar to those observed in the LMC, with IMF slopes $\alpha =
2.35$ (Salpeter 1955; where the IMF is characterised as $\phi(m_\ast)
\propto m_\ast^{-\alpha}$, as a function of stellar mass, $m_\ast$)
and $\alpha = 1.50$, i.e., roughly covering the range of (present-day)
mass function slopes observed in LMC clusters at the time he performed
his {\it N}-body simulations (see also de Grijs et al. 2002b,c). The
stellar mass range covered ranged from 0.15 to $15 M_\odot$; his {\it
N}-body runs spanned at most a few 100 Myr. Following Chernoff \&
Weinberg (1990), and based on a detailed comparison between the
initial conditions for the LMC YMCs derived in Goodwin (1997b) and the
survival chances of massive star clusters in a Milky Way-type
gravitational potential (Goodwin 1997a), Goodwin (1997b; see also
Takahashi \& Portegies Zwart 2000, their fig. 8) concluded that -- for
Galactocentric distances $\ga 12$ kpc -- some of his simulated LMC
YMCs should be capable of surviving for a Hubble time if $\alpha \ge
2$ (or even $\ga 3$; Mengel et al. 2002), but not for shallower IMF
slopes for any reasonable initial conditions (cf. Chernoff \& Shapiro
1987; Chernoff \& Weinberg 1990). More specifically, Chernoff \&
Weinberg (1990) and Takahashi \& Portegies Zwart (2000), based on
numerical cluster simulations employing the Fokker-Planck
approximation, suggest that the most likely survivors to old age are,
additionally, characterised by King model concentrations, $c \ga
1.0-1.5$. Mengel et al. (2002; their fig. 9) use these considerations
to argue that their sample of YMCs observed in the Antennae
interacting system might survive for at least a few Gyr, but see de
Grijs et al. (2005a), and Bastian \& Goodwin (2006) and Goodwin \&
Bastian (2006), for counterarguments related to environmental effects
and to variations in the clusters' SFEs, respectively.

In addition, YMCs are subject to a variety of additional internal and
external drivers of cluster disruption. These include internal
two-body relaxation effects, the nature of the stellar velocity
distribution function, the effects of stellar mass segregation, disk
and bulge shocking, and tidal truncation (e.g., Chernoff \& Shapiro
1987; Gnedin \& Ostriker 1997). All of these act in tandem to
accelerate cluster expansion, thus leading to cluster dissolution --
since expansion will lead to greater vulnerability to tidally-induced
mass loss.

\subsection{Survival Diagnostics: the Mass-to-Light Ratio versus Age 
Diagram}

\begin{figure}[h!]
   \plotone{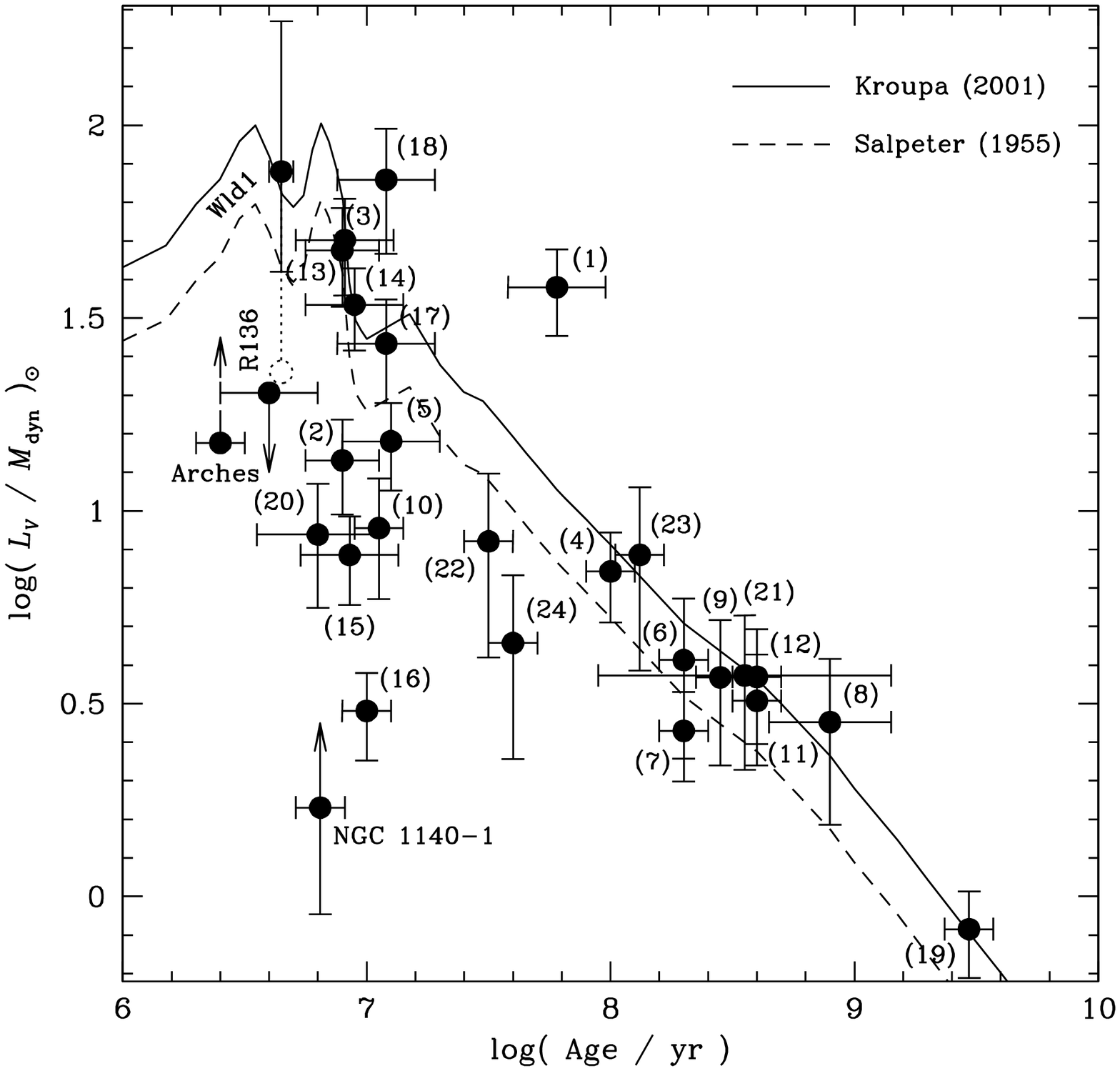}
   \caption{Updated version of the YMC $M/L$ ratio versus age
   diagnostic diagram. The numbered data points were taken from
   Bastian et al. (2006; and references therein); overplotted are the
   SSP predictions for a Salpeter (1955) and a Kroupa (2001) stellar
   IMF. We have included four new YMCs, NGC 1140-1 (Moll et al. 2006)
   the Galactic Centre Arches cluster (see Sect.  \ref{gc.sec}), R136
   in 30 Doradus (Goodwin \& Bastian 2006) and Westerlund 1 (denoted
   `Wld 1'; Sect. \ref{gc.sec}). Regarding the latter, the vertical
   error bars are entirely due to the uncertainty in the dynamical
   mass estimate of Mengel \& Tacconi-Garman (2007); the uncertainty
   introduced by the various foreground extinction estimates is shown
   by the dotted line (see text).\hfill\eject {\it YMC identifications
   --} (1): M82-F; (2): M82 MGG-9; (3): M82 MGG-11; (4): NGC 5236-502;
   (5): NGC 5236-805; (6): NGC 4214-10; (7): NGC 4214-13; (8): NGC
   4449-27; (9): NGC 4449-47; (10): NGC 6946-1447; (11): NGC 7252 W3;
   (12): NGC 7252 W30; (13): (Antennae) W99-1; (14): W99-2; (15):
   W99-15; (16): W99-16; (17): NGC 1569 A1; (18): NGC 1705-I; (19):
   NGC 1316 G114; (20): ESO 338-IG04 \#23; (21): ESO 338-IG04 \#34;
   (22): NGC 1850; (23): NGC 1866; (24): NGC 2157.}
   \label{diagnostic.fig}
\end{figure}

With the ever increasing number of large-aperture ground-based
telescopes equipped with state-of-the-art high-resolution
spectrographs and the wealth of observational data provided by the
{\sl HST}, we may now finally be getting close to resolving the issue
of potential YMC longevity conclusively. To do so, one needs to obtain
(i) high-resolution spectroscopy, in order to obtain dynamical mass
estimates, and (ii) high-resolution imaging to measure their sizes
(and luminosities). As a simple first approach, one could then
construct diagnostic diagrams of YMC mass-to-light ($M/L$) ratio
versus age, and compare the YMC loci in this diagram with simple
stellar population (SSP) models using a variety of IMF descriptions
(cf. Smith \& Gallagher 2001; Mengel et al. 2002; Bastian et al. 2006;
Goodwin \& Bastian 2006). In Fig. \ref{diagnostic.fig} we present an
updated version of the $M/L$ ratio versus age diagram, including all
of the YMCs for which the required observables are presently
available. However, such an approach, while instructive, has serious
shortcomings. The viability of this approach depends, in
essence\footnote{To a small extent, this also depends on our
assumptions of spherical symmetry and the fact that the clusters are
gravitationally bound. While the effect of slight non-sphericity will
lead to underestimates of $\sigma_{\rm los}$, this effect is almost
negligible compared to the other effects described here, except for
M82-F (see Sect. \ref{m82f.sec}). As long as we consider compact
($r_{\rm h} \la$ a few pc) clusters older than a few Myr, the
assumption that they are gravitationally bound is reasonable, since
typical crossing times are $t_c = r_{\rm h} \sigma_{\rm los}^{-1} \sim
1$ Myr (e.g., Larsen 1999, his table 4.4).}, on the validity of the
virial equation to convert line-of-sight velocity dispersions,
$\sigma_{\rm los}^2$, to dynamical mass estimates, $M_{\rm dyn}$, via
(Spitzer 1987):

\begin{equation}
\label{virial.eq}
M_{\rm dyn} = \frac{\eta \sigma^2_{\rm los} r_{\rm h}}{G} \quad ,
\end{equation}
where $r_{\rm h} = 1.3 \; R_{\rm eff}$ are the half-mass and effective
(or half-light) radii of the cluster, respectively, and $\eta = 3 a$;
$a \approx 2.5$ is the factor required to convert the half-mass to the
gravitational radius, $r_{\rm g}$. More specifically, following Fleck
et al. (2006), we write
\begin{equation}
\label{gravrad.eq}
r_{\rm g} = \frac{5}{2} \times \frac{4}{3} r_{\rm h} \quad ,
\end{equation}
where the factor 5/2 provides an approximate conversion for a large
range of clusters characterised by King (1966) mass profiles; the
second numerical factor in Eq. (\ref{gravrad.eq}) results from
projection on the sky, assuming that light traces mass throughout the
cluster. The use of both Eq. (\ref{virial.eq}) and the $M/L$ ratio
versus age diagram rely on a number of assumptions and degeneracies,
however, which we will discuss in detail below.

\subsubsection{IMF degeneracies}

In the simplest approach, in which one compares the YMC loci in the
$M/L$ ratio versus age diagram with SSP models, the data can be
described by {\it both} variations in the IMF slope {\it and}
variations in a possible low-mass cut-off (e.g., Sternberg 1998; Smith
\& Gallagher 2001; Mengel et al. 2002); the models are fundamentally
degenerate for these parameters. For instance, based on $K$-band
observations, Mengel et al. (2002) find that their sample of high-mass
YMCs in the Antennae system contains a subgroup of clusters with
either a relatively shallow IMF slope down to the hydrogen-burning
limit, $m_\ast \sim 0.1 M_\odot$ (the exact value depending on
metallicity), or perhaps with a slightly steeper slope but a truncated
IMF at higher stellar masses, $m_\ast \sim 1 M_\odot$. They note that
these alternatives are less apparent in the $V$ band.

Similarly, Sternberg (1998) derived for the YMC NGC 1705-I that it
must either have a flat mass function ($\alpha < 2$) or a low-mass
truncation between 1 and $3 M_\odot$ (see also Smith \& Gallagher
2001); in both cases, it is unlikely that this cluster may be capable
of surviving for a Hubble time. In addition, Smith \& Gallagher
(2001), using a similar approach, found that the unusual YMC M82-F may
also be characterised by a low-mass cut-off of its mass function at
$\sim 2-3 M_\odot$, or perhaps by a shallow slope, $\alpha \sim 2$ for
a mass function that is well sampled down to the hydrogen-burning
limit (but see Sect. \ref{m82f.sec} for a more in-depth discussion on
this object).

However, the conclusion that the IMFs of these clusters may be unusual
must be regarded with caution. As Smith \& Gallagher (2001) point out,
previous claims for highly abnormal (initial) mass functions have
often proven incorrect. If anything, the shape of the mass function
may vary on the size scales of the individual clusters, but once one
considers their birth environments on larger scales the present-day
mass function appears to be remarkably robust (e.g., Scalo 1998,
Kroupa 2001), with the possible exception of the resolved starburst
clusters in the Milky Way (e.g., Stolte et al. 2005, 2006), NGC 3603
and -- in particular -- the Galactic Centre Arches cluster (see
Sect. \ref{gc.sec}).

Despite this controversy (particularly for some of the youngest
clusters), it appears that most of the YMCs for which high-resolution
spectroscopy is available are characterised by ``standard'' Salpeter
(1955) or Kroupa (2001) IMFs (e.g., Larsen et al. 2001; McCrady,
Gilbert \& Graham 2003; Maraston et al. 2004; Larsen, Brodie \& Hunter
2004; Larsen \& Richtler 2004; Bastian et al. 2006; see also de Grijs
et al. [2005a] for a comparison of dynamical and photometric masses,
the latter based on ``standard'' IMF representations). In addition,
significant recent progress has been made in our understanding of the
properties of the YMCs characterised by apparently non-standard IMFs,
which we will discuss in detail in Sect. \ref{expulsion.sec}.

\subsubsection{Mass segregation}

While the assumption that these objects are {\it approximately} in
virial equilibrium is probably justified at ages greater than a few
$\times 10^7$ yr and for realistic SFEs $\ga 30$ per cent (at least
for the stars dominating the light; see, e.g., Goodwin \& Bastian
2006), the {\it central} velocity dispersion (as derived from
luminosity-weighted high-resolution spectroscopy) does not necessarily
represent a YMC's total mass. It is now well-established that almost
every YMC exhibits significant mass segregation from very young ages
onwards, so that the effects of mass segregation must be taken into
account when converting central velocity dispersions into dynamical
mass estimates (see also Lamers et al. 2006; Fleck et al. 2006; Moll
et al. 2006; S. L. Moll et al., in prep.).

By ignoring the effects of mass segregation, as is in essence done if
one simply applies Eq. (\ref{virial.eq}), the underlying assumption is
then that of an isotropic stellar velocity distribution, i.e.,
$\sigma^2_{\rm total} = 3 \sigma^2_{\rm los}$, where $\sigma^2_{\rm
total}$ is the cluster's mean three-dimensional velocity dispersion.
In the presence of (significant) mass segregation in a cluster, the
central velocity dispersion will be dominated by the higher-mass stars
populating the cluster core. If we focus on dynamical evolution as the
dominant cause of mass segregation in clusters (as opposed to the
possible preferential formation of the higher-mass stars close to the
cluster core, also known as ``primordial'' mass segregation; e.g.,
Bonnell \& Davies 1998; de Grijs et al. 2002b), it follows that for
the high-mass stars to migrate to the cluster core, i.e., to the
bottom of the gravitational potential, they must have exchanged some
of their kinetic energy with their lower-mass counterparts on more
extended orbits. As a consequence, the velocity dispersion dominating
the observed high-resolution spectra will be {\it lower} than expected
for a non-mass-segregated cluster of the same mass. In addition,
measurements of $r_{\rm h}$ will also be biased to smaller values, and
not to the values associated with the cluster as a whole. Mass
segregation will thus lead to an {\it under}estimate of the true
cluster mass.

We also note that the assumption of virial equilibrium only holds to a
limited extent, even in old GCs, because cluster-wide relaxation
time-scales of massive GC-type objects are of order $10^9$ yr or
longer (Djorgovski 1993). In fact, full global, or even local, energy
equipartition among stars covering a range of masses is never reached
in a realistic star cluster, not even among the most massive species
(e.g., Inagaki \& Saslaw 1985; Hunter et al. 1995). As the dynamical
evolution of a cluster progresses, low-mass stars will, on average,
attain larger orbits than the cluster's higher-mass stars, and the
low-mass stars will thus spend most of their time in the cluster's
outer regions, at the extremes of their orbits. For this reason alone,
we would not expect to achieve global energy equipartition in a
cluster. 

The time-scale for the onset of significant dynamical mass segregation
is comparable to the cluster's dynamical relaxation time (Spitzer \&
Shull 1975; Inagaki \& Saslaw 1985; Bonnell \& Davies 1998; Elson et al. 
1998). A cluster's characteristic time-scale may be taken to be its
half-mass (or median) relaxation time, i.e., the relaxation time at the
mean density for the inner half of the cluster mass for cluster stars
with stellar velocity dispersions characteristic for the cluster as a
whole (Spitzer \& Hart 1971; Lightman \& Shapiro 1978; Meylan 1987;
Malumuth \& Heap 1994).

Although the half-mass relaxation time characterises the dynamical
evolution of a cluster as a whole, significant differences are
expected locally within the cluster. The relaxation time-scale will be
shorter for higher-mass stars than for their lower-mass companions;
numerical simulations of realistic clusters confirm this picture
(e.g., Aarseth \& Heggie 1998; Kim et al. 2000; Portegies Zwart et
al. 2002). From this argument it follows that dynamical mass
segregation will also be most rapid where the local relaxation time is
shortest, i.e., near the cluster centre (cf. Fischer et al. 1998,
Hillenbrand \& Hartmann 1998). Thus, significant mass segregation
among the most massive stars in the cluster core occurs on the local,
central relaxation time-scale (comparable to just a few crossing
times; cf. Bonnell \& Davies 1998).

The combination of these effects will lead to an increase of the
dimensionless parameter $\eta$ in Eq. (\ref{virial.eq}) with time, if
the characteristic two-body relaxation time of a given (massive)
stellar species is short (Boily et al. 2005; Fleck et al. 2006), and
thus to an {\it under}estimate of the true cluster mass. However, we
note that Goodwin \& Bastian (2006) point out that a large fraction of
the youngest clusters in the $M/L$ ratio versus age diagram appear to
have dynamical masses well in excess of their photometric masses, and
that, therefore, the result of Boily et al. (2005) and Fleck et
al. (2006) does not seem applicable to these YMCs.

\subsubsection{Stellar masses}

Estimating dynamical masses via Eq. (\ref{virial.eq}) assumes, in
essence, that all stars in the cluster are of equal mass. This is
clearly a gross oversimplification, which has serious consequences for
the resulting mass estimates. The straightforward application of the
virial theorem tends to {\it under}estimate a system's dynamical mass
by a factor of $\sim 2$ compared to more realistic multi-mass models
(e.g., Mandushev et al. 1991; based on an analysis of the
observational uncertainties). Specifically, Mandushev et al. (1991)
find that the mass-luminosity relation for GCs with mass
determinations based on multi-component King-Michie models (obtained
from the literature) lies parallel to that for single-mass King
models, but offset by $\Delta \log(m_{\rm cl} [M_\odot]) \simeq 0.3$
towards higher masses. Farouki \& Salpeter (1982) already pointed out
that cluster relaxation and its tendency towards stellar energy
equipartition is accelerated as the stellar mass spectrum is widened;
mass segregation will then take place on shorter time-scales than for
single-component (equal mass) clusters, and thus this will once again
lead to an {\it under}estimate of the true cluster mass (see also
Goodwin 1997a; Boily et al. 2005; Fleck et al. 2006 for multi-mass
{\it N}-body approaches).

We also point out that if the cluster contains a significant fraction
of primordial binary and multiple systems, these will act to
effectively broaden the mass range and thus also speed up the
dynamical evolution of the cluster (e.g., Fleck et al. 2006).

\subsubsection{Can We Reduce the Number of Assumptions?}

Motivated by a desire to reduce the number of assumptions involved in
these dynamical mass estimates, in de Grijs et al. (2005a) we explored
whether we could constrain the evolution of individual YMCs based on
their loci in the plane defined by their luminosities (or absolute
$V$-band magnitudes, $M_V$) and central velocity dispersions,
$\sigma_0$. The method hinges on the empirical relationship for old
GCs in the Local Group, which occupy a tightly constrained locus in
this plane (Djorgovski et al. 1997; McLaughlin 2000; and references
therein). We concluded that the tightness of the relationship for a
sample drawn from environments as diverse as those found in the Local
Group, ranging from high to very low ambient densities, implies that
its origin must be sought in intrinsic properties of the GC formation
process itself, rather than in external factors.

Encouraged by the tightness of the GC relationship, we also added the
available data points for the YMCs in the local Universe for which
velocity dispersion information was readily available. In order to
compare them to the old Local Group GCs, we evolved their luminosities
to a common age of 12 Gyr, adopting the ``standard'' Kroupa (2001) IMF
(for stellar masses from 0.1 to $100 M_\odot$). We found that the
($\sim \frac{2}{3}$) majority of our sample YMCs end up scattering
closely about the Local Group GC relationship. In the absence of
significant external disturbances, this would imply that these objects
may potentially survive to become old GC-type objects by the time they
reach a similar age. In order to investigate whether dynamical
evolution would have a dramatic impact on the evolution of clusters in
the $M_V - \sigma_0$ plane, we analysed the results of a number of
{\it N}-body simulations. We showed that the evolution of the observed
$\sigma_0$ is relatively small for clusters that survive to old age,
and in the sense that the central velocity dispersion decreases with
time.

Thus, these results provide additional support to the suggestion that
the formation of proto-GCs appears to be continuing until the
present. Detailed case by case comparisons between our results in de
Grijs et al. (2005a) with those obtained previously and independently
based on dynamical mass estimates and $M/L$ ratio considerations lend
significant support to the feasibility and robustness of this novel
method (see de Grijs et al. 2005a for a detailed discussion). The main
advantage of this method compared to the more complex analysis
involved in using dynamical mass estimates is its simplicity and
empirical basis. The only observables required are the system's
line-of-sight velocity dispersion and photometric properties.

\subsection{Gas Expulsion}
\label{expulsion.sec}

It may appear that a fair fraction of the $\sim 10$ Myr-old YMCs that
have been analysed thus far might be characterised by unusual IMFs,
since their loci in the diagnostic diagram are far removed from any of
the ``standard'' SSP models (see, e.g., Smith \& Gallagher 2001;
Bastian et al. 2006; Goodwin \& Bastian 2006; Moll et al. 2006; S. L.
Moll et al., in prep.). However, Bastian \& Goodwin (2006) and Goodwin
\& Bastian (2006) recently showed that this is most likely an effect
of the fact that the velocity dispersions of these young objects do
not adequately trace their masses. Using a combination of
high-resolution {\sl HST} observations of the YMCs M82-F, NGC 1569-A
and NGC 1705-I combined with {\it N}-body simulations of clusters that
have recently undergone the rapid removal of a significant fraction of
their mass due to gas expulsion (e.g., caused by supernova activity
and massive stellar winds), Bastian \& Goodwin (2006) suggest that
these clusters are undergoing violent relaxation and are also subject
to {\it stellar} mass loss. As a result, the observed velocity
dispersions of these YMCs will include the non-gravitational motions
of the escaping stars, and thus lead to artificially enhanced
dynamical mass estimates. More specifically, Bastian \& Goodwin (2006)
derive that such dynamical mass estimates may be wrong by a factor of
up to 3 for 10--20 Myr after gas expulsion, for SFEs $\ga 40$ per
cent, as for YMCs of similar age as NGC 1569-A and NGC 1705-I, as well
as for many young LMC clusters.

Although this discrepancy between the dynamical and the true stellar
mass will disappear quickly (i.e., within 10--15 Myr) for clusters
with SFEs $\ga 50$ per cent, clusters with lower SFEs are more
significantly affected. Bastian \& Goodwin (2006) show that for YMCs
with SFE $\sim 40$ per cent, the apparent dynamical mass will be much
greater than the true stellar mass for YMCs at an age of $\sim 10$
Myr, and virial equilibrium is not achieved until an age of $\sim 50$
Myr. The observed stellar velocity dispersions of such clusters may
even {\it under}estimate the true stellar mass by up to 30 per cent,
as the cluster has overexpanded in its attempt to reach a new virial
equilibrium.

Thus, claims of abnormal stellar (initial) mass functions based on
such erroneous dynamical mass estimates, which would clearly affect
the clusters' loci in the $M/L$ ratio versus age diagram, should be
treated with extreme caution (cf. de Grijs et al. 2005a; Bastian et
al. 2006; Bastian \& Goodwin 2006; Goodwin \& Bastian 2006). In this
respect, it is encouraging to see that the older clusters (i.e., older
than a few $\times 10^7 - 10^8$ yr) seem to conform to ``normal''
IMFs; by those ages, the clusters' velocity dispersions seem to
represent the underlying gravitational potential much more closely;
see Goodwin \& Bastian (2006; see also Fig. \ref{diagnostic.fig}) for
the most recently updated $M/L$ ratio versus age diagram. Goodwin \&
Bastian (2006) show indeed that, depending on the cluster's SFE, an
equilibrium state is reached after $\sim 30$ Myr, when the effects of
gas expulsion are no longer easily detectable in the cluster's stellar
velocity dispersion (at least for SFEs $\ga 40$ per cent; lower-SFE
clusters most likely disperse into the background field star
population well before that age). In fact, Goodwin \& Bastian (2006)
show conclusively that the scatter in the YMC loci for the youngest
ages ($\sim 10^7$ yr) in this diagnostic diagram can be fully
understood if we vary the effective SFE to allow for SFEs as low as
about 10 per cent. This then implies that those YMCs that lie well
below the diagnostic line based on the ``standard'' IMF evolution will
disperse; these objects include, among others, many of the YMCs in the
Antennae galaxies (Mengel et al. 2002; see also de Grijs et al. 2005a
for a discussion of these YMCs), some of the young LMC clusters
(McLaughlin \& van der Marel 2005), including R136 in 30 Doradus (see
Goodwin \& Bastian 2006), NGC 1569-A and NGC 1705-I (e.g., Smith \&
Gallagher 2001), NGC 6946-1447 (Larsen et al. 2006; corrected for
extinction in Goodwin \& Bastian 2006), YMC number 23 in ESO 338-IG04
(\"Ostlin, Cumming \& Bergvall 2007), and NGC 1140-1 (Moll et al.
2006). Of all these discrepant YMCs, NGC 1140-1 appears to have the
largest velocity dispersion (and hence dynamical mass estimate). This
is most likely due to its environmental effects; the YMC is situated
in a vigorously star-forming region of its host galaxy, and is
possibly interacting with a neighbouring YMC (de Grijs et al. 2004;
S. L. Moll et al., in prep.). In addition, the ground-based
high-resolution spectra likely include flux from the nearest
neighbouring clusters as well, which might enhance the observed
velocity dispersion of the region covered. As such, one should treat
this data point with due caution.

\subsection{The Enigmatic Cluster M82-F}
\label{m82f.sec}

Thanks to the recent studies of Bastian \& Goodwin (2006) and Goodwin
\& Bastian (2006), we are now confident that we understand why some of
the YMCs deviate significantly from the expected evolution -- for a
``standard'' IMF -- of the stellar $M/L$ ratio at ages of $\sim 10^7$
yr, as discussed in the previous section. However, the location of the
massive cluster F in M82 remains to be understood, despite having been
the subject of a considerable number of recent studies (e.g.,
O'Connell et al. 1995; Gallagher \& Smith 1999; Smith \& Gallagher
2001; McCrady et al. 2003, 2005; Bastian \& Goodwin 2006; Bastian et
al. 2007).

Smith \& Gallagher (2001) concluded that, assuming that their age ($60
\pm 20$ Myr), mass ($1.2 \pm 0.1 \times 10^6 M_\odot$) and luminosity
($M_V = -14.5 \pm 0.3$ mag; $L_V / M_{\rm dyn} = 45 \pm 13 \;
L_{V,\odot} / M_\odot$) measurements for the object were correct, it
must either have a low-mass cut-off of its mass function at $\sim 2-3
M_\odot$, or a shallow mass function slope, $\alpha \sim 2$ for a mass
function including stellar masses down to $0.1 M_\odot$. McCrady et
al. (2003) suggested, based on near-infrared population synthesis
modeling, that Smith \& Gallagher (2001) had overestimated their ages
as well as the half-mass radius of the cluster; they derived the
latter more accurately based on higher-resolution ({\sl HST}/ACS)
imaging than available to Smith \& Gallagher (2001). However, in a
recent contribution, Bastian et al. (2007) used optical spectroscopy,
in essence confirming the YMC's age to be in the range from 50--70
Myr. However, contrary to the earlier studies, Bastian et al. (2007)
conclude that M82-F is subject to a large amount of differential
extinction, thus rendering earlier luminosity estimates based on
foreground-screen extinction very uncertain. They conclusively show
that the apparently large degree of mass segregation in the cluster,
derived by McCrady et al. (2005) based on the variation of $r_{\rm h}$
as a function of wavelength, is in fact caused by this differential
extinction. Applying their spatially resolved extinction corrections
to the {\sl HST}/ACS data, Bastian et al. (2007) also confirm the size
estimate of Smith \& Gallagher (2001). Thus, Bastian et al. (2007)
have firmly placed the M82-F data point close to its original location
well away from the prediction for the evolution of an SSP
characterised by a ``standard'' IMF in the diagnostic $M/L$ ratio
versus age diagram.

Smith \& Gallagher (2001), Bastian et al. (2006) and Bastian \&
Goodwin (2006) have all suggested that the discrepancy between the
observations of M82-F and the evolution of an SSP with a ``standard''
IMF could be resolved if the YMC were as young as $\sim 15$
Myr. However, while the cluster's structure reveals evidence for
non-virial equilibrium at large radii (Bastian \& Goodwin 2006),
non-sphericity (Smith \& Gallagher 2001; McCrady et al. 2005), clumpy
substructure (e.g., Bastian et al. 2007) and multiple stellar
populations (see Bastian et al. 2007 for details), none of these
properties seem capable of resolving the age issue; the current best
age estimate for M82-F remains close to 60 Myr (Bastian et al. 2007).
In fact, based on its location in the diagnostic diagram, the object
appears to be subvirial instead of supervirial as expected if the
effects of gas expulsion were significant. In addition, any subvirial
phase would be expected to occur shortly after the main gas expulsion
phase, when the cluster is attempting to regain virial equilibrium;
the current best YMC age estimate is too old for this to be a likely
evolutionary state for M82-F, however. Secondly, the young age
solution seems to be ruled out by the spectroscopic data of Bastian et
al. (2007); we note that the cluster is projected onto a background
H{\sc ii} region (e.g., Smith \& Gallagher 2001), which might affect
the integrated photometry.

Thus, we conclude that the YMC M82-F remains an enigma. Based on the
simplest assumptions it appears that its present-day mass function may
be deficient in low-mass stars, although its complicated substructure
might imply that there are unknown factors at work here (such as tidal
disruption in the intense tidal field in the M82 disk). If the
cluster's mass function is indeed ``top heavy'', it is unlikely to
survive for more than 1--2 Gyr (Smith \& Gallagher 2001; McCrady et
al. 2005).

\subsection{Galactic starburst clusters}
\label{gc.sec}

Thus far, the best evidence yet for an unusual present-day mass
function in a compact, massive young star cluster relates to the $\sim
2$ Myr-old Arches cluster near the Galactic Centre. Stolte et
al. (2005) present direct evidence for a low-mass depleted mass
function, based on a turn-over of the cluster's mass function around
6--$7 M_\odot$, i.e., well above their observational detection limit
in the cluster core (at $\sim 4 M_\odot$) decreasing to $\sim 2
M_\odot$ at large radii. They hypothesize that the significant
overdensity of high-mass stars and the depletion of low and
intermediate-mass stars in the cluster core could be caused by either
rapid dynamical mass segregation or the preferential formation of
high-mass stars with respect to their lower-mass counterparts (i.e.,
primordial mass segregation; see also Figer et al. 1999), while at
larger radii they suggest that either tidal stripping of lower-mass
stars ($m_\ast \la 6 M_\odot$) by the Galactic Centre's tidal field
(based on the rapid dissolution time-scales suggested by {\it N}-body
simulations; Kim et al. 2000; Portegies Zwart et al. 2002), or a
physically truncated {\it initial} mass function might be responsible.
In the latter case, this might imply that a low-mass truncated and
possibly (Figer et al. 1999; Stolte et al. 2002, 2005) also a somewhat
flattened IMF may be more appropriate for starburst YMCs\footnote{A
similar flattening of the IMF slope has been found for the young
starburst cluster R136 in 30 Doradus in the LMC (Sirianni et al.
2000), where the slope flattens below $m_\ast \simeq 2 M_\odot$, and
in the embedded young Galactic (starburst) cluster NGC 2024, which
shows a flattening of the IMF slope below $m_\ast \sim 1 M_\odot$
(Comeron, Rieke \& Rieke 1996).}. If so, this will have significant
consequences for stellar population synthesis modeling of
extragalactic starburst environments and the derivation of integrated
properties. An apparent flattening of the IMF was also observed in NGC
3603 (Stolte et al. 2006; but see Eisenhauer et al. 1998) although
this YMC's mass function does not appear to be truncated at low masses
(Eisenhauer et al. 1998; Sung \& Bessell 2004; Stolte et al. 2006).

For completeness, we have added the appropriate data point for the
Arches cluster to the diagnostic $M/L$ ratio versus age diagram of
Fig. \ref{diagnostic.fig}. Figer et al. (2002) derive an age of $2.5
\pm 0.5$ Myr, and an upper limit to the mass of the Arches cluster of
$m_{\rm cl} < 7 \times 10^4 M_\odot$ within a radius of 0.23 pc, which
they claim to be ``about 5 times greater than what would be expected
from the mass function'' in Figer et al. (1999). The latter,
photometric, mass estimate of 10,800 $M_\odot$ is based on
extrapolating the YMC's (significantly flattened) IMF to a lower mass
limit of $1 M_\odot$, although adopting a lower mass limit of $0.1
M_\odot$ results in a total mass that is merely a factor of $\approx
1.1$ greater. Thus, the Arches data point in Fig. \ref{diagnostic.fig}
is located at a $\sim 5$ times smaller light-to-mass ratio than
expected for a Kroupa-type IMF (which we deem to be a more appropriate
IMF for the purpose of this comparison than the single power-law
Salpeter IMF). Although the YMC's location in Fig.
\ref{diagnostic.fig} suggests that its SFE may be as low as 30 per
cent (cf. fig. 5 in Goodwin \& Bastian 2006), we point out that (i)
our value is in fact a lower limit, and (ii) its location near the
Galactic Centre implies that it is subject to significant external
pressures, which are likely to inject additional energy into the
system, thus artificially increasing its velocity dispersion.
Nevertheless, whether because of the internal dynamics, or because of
external tidal effects, given its current mass function the Arches
cluster is unlikely to survive for a Hubble time.

Possibly the best local example of a YMC in the Galaxy is Westerlund 1
($m_{\rm cl} \sim 10^5 M_\odot$, with an absolute lower limit of
$m_{\rm cl,low} \simeq 1.5 \times 10^3 M_\odot$; Clark et al. 2005;
see also Mengel \& Tacconi-Garman 2007, who recently determined
$m_{\rm cl} = 6.3^{+5.3}_{-3.7} \times 10^4 M_\odot$, based on its
near-infrared velocity dispersion and assuming the object to be in
virial equilibrium), at an age of 4--5 Myr (Crowther et al. 2006). Due
to the large extinction\footnote{Clark et al. (2005) derive an updated
value of $A_V = 11.6$ mag. We include Westerlund 1 in Fig.
\ref{diagnostic.fig}, adopting the recent dynamical mass estimate of
Mengel \& Tacconi-Garman (2007), the integrated $V$-band luminosity
out to a radius $R \simeq 1.4$ pc derived from fig. 8 of Piatti et
al. (1998; corrected for a current best distance estimate of $D = 5.5$
kpc; Clark et al. 2005), and the current best age estimate of Crowther
et al. (2006). We also show the object's location in this diagram if
we had assumed Clark et al.'s (2005) extinction estimate (dotted
line).} towards the cluster, $A_V \sim 12.9$ mag (Piatti et al. 1998),
its stellar mass function below $m_\ast \sim 1.5 M_\odot$ is
essentially unconstrained by direct observations (e.g., Clark et
al. 2007; and references therein). Clark et al.'s (2005) mass estimate
of $m_{\rm cl} \sim 10^5 M_\odot$ was based on the assumption of a
Kroupa (2001) stellar IMF. By comparison of the X-ray fluxes of the
Orion Nebula Cluster (ONC) and Westerlund 1, Clark et al. (2007)
speculate that the 45 X-ray bright pre-main sequence (pre-MS)
candidate stars in Westerlund 1 form the high-luminosity tail of $\ge
36,000$ lower-luminosity (and hence lower-mass, $m_\ast < 1.5
M_\odot$) pre-MS stars.

On the other hand, Muno et al. (2006) argue that the non-thermal
spectrum of the diffuse X-ray emission from Westerlund 1 and the ONC
are markedly (physically) different, in the sense that the stellar
metallicities are much lower (around solar) in Westerlund 1 than in
the ONC. This implies (Muno et al. 2006) that the low-mass stars ($0.3
M_\odot < m_\ast < 2 M_\odot$) contribute $\la 30$ per cent of the
diffuse X-ray flux from Westerlund 1, corresponding to $\la 40,000$
low-mass stars in this mass range. If the YMC's stellar IMF were of
the Kroupa (2001) type, one would expect $\ga 100,000$ stars in this
mass range. This result seems to imply, therefore, that either the
stellar IMF in Westerlund 1 is flat ($\alpha \la 2.1$), or that it is
truncated at $m_\ast < 0.6 M_\odot$ (Muno et al. 2006). In the latter
case, the total mass of the cluster would not change significantly
under the assumption of a Kroupa (2001) IMF; in the former case, the
total mass of the cluster would only be $m_{\rm cl} \sim 40,000 -
70,000 M_\odot$. In either case, it is unlikely that Westerlund 1 will
survive for a Hubble time. However, while the Muno et al. (2006) and
the Clark et al. (2007) estimates seem to constrain the number of
low-mass stars in the cluster quite well, we note that the arguments
presented here are based on observations of the massive stars only,
and hence on assumed extrapolations of the mass function to lower
masses. These results must therefore be treated with due caution.

\subsection{A ``Super'' Star Cluster Grown Old?}

We recently reported the discovery of an extremely massive, but old
($12.4 \pm 3.2$ Gyr) GC in M31, 037-B327, that has all the
characteristics of having been an exemplary YMC at earlier times,
based on an extrapolation of its present-day extinction-corrected
$V$-band luminosity back to an age of 10 Myr (Ma et al. 2006b; see also
Cohen 2006). To have survived for a Hubble time, we conclude that its
stellar IMF cannot have been top-heavy. Using this constraint, and a
variety of SSP models, we determined a {\it photometric} mass for
037-B327 of $M_{\rm GC} = (3.0 \pm 0.5)\times 10^7 M_\odot$, somewhat
depending on the SSP models used, the metallicity and age adopted and
the IMF representation. In view of the large number of free
parameters, the uncertainty in our photometric mass estimate is
surprisingly small (although this was recently challenged by Cohen
2006). This mass, and its relatively small uncertainties, make this
object potentially one of the most massive star clusters of any age in
the Local Group. As a surviving ``super'' star cluster, this object is
of prime importance for theories aimed at describing massive star
cluster evolution.

Cohen (2006) obtained an optical velocity dispersion for the cluster
using the Keck/HIRES spectrograph, and showed that it is comparable to
that of M31 G1. Depending on the wavelength range used, they find
$\sigma_{\rm los} = 19.2 \pm 3.5$ km s$^{-1}$ ($5150 - 5190$ \AA) and
$\sigma_{\rm los} = 19.9 \pm 3.4$ km s$^{-1}$ ($6545 - 6600$ \AA),
compared to $\sigma_{\rm los} \sim 22$ km s$^{-1}$ for M31 G1. The
cluster's half-light radius, $r_{\rm h} \simeq 2.5 \pm 0.2$ pc (Ma et
al. 2006b). For M31 G1, we recently redetermined $r_{\rm h} = 6.5 \pm
0.3$ pc (Ma et al. 2007). Thus, based on these most recent results, it
appears that 037-B327 may be a factor of $\sim 2-3$ less massive than
M31 G1, assuming that both GCs have the same stellar IMF.
Nevertheless, this still confirms the nature of 037-B327 as one of the
most massive star clusters in the Local Group.

Cohen (2006) suggests that the high mass estimate of Ma et al. (2006b)
may have been affected by a non-uniform extinction distribution across
the face of the cluster (see also Ma et al. 2006a for a more detailed
discussion). She obtains, from new $K$-band imaging and different
assumptions on the extinction affecting the $K$-band light, that $M_K$
of 037-B327 may be some 0.16 mag brighter than that of M31 G1, or
about twice as luminous. Despite these corrections provided by Cohen
(2006), the basic conclusion from Ma et al. (2006b), i.e., that at the
young age of 10 Myr cluster M31 037-B327 must have been a benchmark
example of a ``super'' star cluster, and that its IMF must thus have
contained a significant fraction of low-mass stars, still stands
firmly.

In this context, it is interesting to note that we recently also
determined photometric masses for the YMCs in the interacting system
NGC 6745, the ``Bird's Head Galaxy'' (de Grijs et al. 2003c). NGC 6745
contains a significant population of luminous (and therefore
presumably massive) ``super'' star clusters. Using the stellar
population synthesis method developed in Anders et al. (2003; see also
de Grijs et al. 2003b), with which we obtained robust, independent
estimates of the star cluster ages, metallicities and extinction from
the shape of their broad-band spectral energy distributions (SEDs),
and masses from a simple scaling of our model SEDs to the
observations, we concluded that a number of the most massive star
clusters in this galaxy are characterised by masses in the range $6.5
< \log( m_{\rm cl} [M_\odot] ) < 8.3$.

These surprisingly high masses are much larger than those of the most
massive GCs in the Milky Way or other galaxies in the Local Group.
However, the mass determination via population synthesis models is
affected by uncertainties in the age and metallicity derivation. 

For the two highest-mass clusters in NGC 6745, we derive a combined
photometric mass of $M_{\rm phot} \sim 6 \times 10^8 M_\odot$. We
should keep in mind, of course, that this high mass estimate is a
strong function of the (low) metallicity assumed; if we had assumed
solar metallicity for these objects, the derived age would have been
significantly smaller ($\sim 10-20$ Myr versus $\sim 1$ Gyr), and the
mass could be smaller by a factor of $\sim 10$. Even so, if we can
verify (S. L. Moll et al., in preparation) our photometric mass
estimates of the two most massive NGC 6745 clusters via spectral
linewidth analysis, these clusters would be the most massive star
clusters known to date, exceeding even the mass of the 300 Myr-old
cluster W3 in the merger remnant galaxy NGC 7252, $M_{\rm dyn} = (8
\pm 2) \times 10^7 M_\odot$ (e.g., Schweizer \& Seitzer 1998; Maraston
et al. 2001, 2004). The latter mass is in excellent agreement with
their photometric mass determination ($M_{\rm phot} \sim 7.2 \times
10^7 M_\odot$; Maraston et al. 2001), assuming a Salpeter stellar IMF.
Thus, in the absence of significant external disturbations, W3 has the
potential to survive for a Hubble time, by virtue of its large
complement of low-mass stars.

The immediate implication of a similar result for the NGC 6745
clusters would be that galaxy interactions indeed produce extremely
massive star clusters, as also suggested by Maraston et al. (2004)
based on their analysis of NGC 7252-W3; see also Pasquali, de Grijs \&
Gallagher (2003) and Pollack, Max \& Schneider (2007) for tentative
indications that the more massive clusters in the galaxy merger NGC
6240 tend to form closer to the most intense interaction regions near
the galaxy's double nucleus. This gives important insights into the
still largely unknown star and star cluster formation processes in
extreme environments.

\section{The Evolution of Star Cluster Systems}
\label{clfs.sec}

Following the violent relaxation induced by the supernova-driven
expulsion of the left-over star-forming gas, star clusters -- at least
those that survive the infant mortality phase -- settle back into
virial equilibrium by the time they reach an age of about 40--50 Myr
(Bastian \& Goodwin 2006; Goodwin \& Bastian 2006). Subsequently, the
initial conditions characterising these gas-free bound star clusters
are modified as secular evolution proceeds. Internal (two-body
relaxation) and external effects (due to interactions with the tidal
field associated with the underlying galactic gravitational potential)
lead to tidal stripping and the evaporation of a fraction of the
low-mass cluster stars, thus resulting in the gradual dissolution of
star clusters (Meylan \& Heggie 1997; Vesperini \& Heggie 1997;
Portegies Zwart et al. 1998; Baumgardt \& Makino 2003; see also
Odenkirchen et al. 2001 and Dehnen et al. 2004 for a study of the
currently dissolving Galactic GC Pal 5).

One of the most important diagnostics used to infer the formation
history, and to follow the evolution of a star cluster population is
the CMF (i.e., the number of clusters per constant logarithmic cluster
mass interval, ${\rm d}N/{\rm d}\log m_{\rm cl}$).\footnote{We adopt
the nomenclature of McLaughlin \& Pudritz (1996). We refer to the
number of objects per {\it linear} mass interval ${\rm d}N/{\rm
d}m_{\rm cl}$ as the mass {\it spectrum}. Where we refer to the mass
{\it function}, this describes the number of objects per {\it
logarithmic} mass interval, ${\rm d}N/{\rm d}\log m_{\rm cl}$.} Of
particular importance is the {\it initial} cluster mass function
(ICMF), since this holds clues to the star and cluster formation
processes. The debate regarding the shape of the ICMF, and of the CMF
in general, is presently very much alive, both observationally and
theoretically. This is so because it bears on the very essence of the
star-forming processes, as well as on the formation, assembly history
and evolution of the clusters' host galaxies over cosmic time. Yet,
the observable property one has access to is the CLF (i.e., the number
of objects per unit magnitude, ${\rm d}N/{\rm d}M_V$). While the
ubiquitous old GCs show a well-established Gaussian CLF, an ever
increasing amount of data suggests that the CLF of YMCs formed in
starbursts and mergering galaxies is best represented by a power-law
function.

\subsection{The CLF of massive star clusters}
\label{subsec:CLF_facts}

\subsubsection{Old Globular Clusters}

Associated with the vast majority of large galaxies, as well as with
the most luminous dwarf galaxies, old GCs are ubiquitous in the local
Universe (e.g., Brodie \& Strader 2006). With ages comparable to that
of the Universe ($\simeq 13$ Gyr), they are regarded as the present
end point of massive compact star cluster evolution. The postulated
evolutionary connection between the old GCs and the massive star
clusters recently formed in interacting and merging galaxies remains a
contentious issue, however. Specifically, the key issue of relevance
here is whether the power-law CLF of YMC populations will evolve into
a bell shape similar to that of the CLF of old GCs.
  
Typical parameters for the Gaussian shape of the GC luminosity
function (GCLF) in the $V$ band are a peak (or ``turn-over'') at a
magnitude of $M_V \simeq -7.4$ mag, and a standard deviation
$\sigma_M$ of $1.2-1.4$ mag (e.g., Harris 1991, 2001; Harris, Harris
\& McLaughlin 1998; Barmby, Huchra \& Brodie 2001; Richtler 2003).
Intriguingly, this shape and parameters seem to be almost universal
among galaxies as they show only a weak dependence on the size, the
morphological type, the metallicity and the environment of the host
galaxy (Ashman, Conti \& Zepf 1995; Kavelaars \& Hanes 1997; Ashman \&
Zepf 1998; Harris 1999; Whitmore et al. 2002). The relative robustness
of the cluster luminosity at the turn-over magnitude has prompted its
use as a distance indicator (Harris 2001; Richtler 2003; but see
Fritze--v. Alvensleben 2004 for a critical assessment), although its
use in dwarf galaxies remains a matter of debate (see Brodie \&
Strader 1996; and references therein). Limited, albeit systematic,
differences in the detailed shape of the GCLF from one galaxy to
another do exist, however (Harris 2001; Brodie \& Strader 2006), such
as, for instance, variations in the extension of the high-luminosity
wing (McLaughlin \& Pudritz 1996; Burkert \& Smith 2000; Jordan et
al. 2006; see also section 3.2 in Parmentier \& Gilmore 2007).

The GCLF constitutes a faithful mirror of the underlying GCMF only if
cluster-to-cluster variations of the integrated cluster $M/L$ ratios
are small. This is true for any cluster system characterised by a
stellar IMF that is invariant, and if the cluster age range is a
limited fraction of the mean age of the cluster population. As regards
the old GCs in the Galactic halo, the range spanned by their visual
$M/L$ ratio is limited (i.e., $1\la M/L_V \la 4$, with a mean $\langle
M/L_V \rangle \sim 2$; Pryor \& Meylan 1993; see also Parmentier \&
Gilmore 2001, their fig. 1), and the scatter partly reflects
variations in the dynamical evolution of individual GCs. It is widely,
and probably safely, assumed that the GCLF is a good proxy of the GCMF
for (old) extragalactic GC systems as well.  Assuming a near-invariant
$M/L$ ratio, $M/L_V \sim 2$, that {\it universal} GCMF is
well-represented by a Gaussian distribution with a mean $\langle \log
(m_{\rm cl}[M_\odot]) \rangle \sim 5.2$--5.3 and a standard deviation
$\sigma _{\log m_{\rm cl}} \sim 0.5$--0.6 dex. The origin of this
universality among galaxies remains an outstanding issue in modern
astrophysics. The peaked shape of the GCMF has sometimes been
interpreted as evidence that GC masses bracket one characteristic
value of $m_{\rm GC} \sim 2 \times 10^5 M_\odot$ (e.g., Peebles \&
Dicke 1968). This is, however, mainly an artefact produced by the
logarithmic binning of the data. The underlying mass spectrum is well
described by a two-index power-law, with exponents $\alpha \sim -2$
and $\alpha \sim -0.2$ above and below $m_{\rm GC} \sim 2 \times 10^5
M_\odot$, respectively (Surdin 1979; McLaughlin 1994). Such a
monotonic behaviour highlights the absence of any GC mass scale.

\subsubsection{Young Massive Star Clusters}

The advent of the {\sl HST} and the subsequent discovery (in
particular in interacting and merging galaxies) of star clusters with
the high luminosities and the compact sizes expected for GCs at young
ages has prompted renewed interest in the evolution of the CLF (and
CMF) of massive star clusters, both observationally and
theoretically. Starting with the seminal work by Elson \& Fall (1985)
on the young LMC cluster system (with ages $\la 2 \times 10^9$ yr), an
ever increasing body of evidence, mostly obtained with the {\sl HST},
seems to imply that the CLF of these YMCs is well described by a power
law of the form ${\rm d}N \propto L^{1+{\alpha}} {\rm d}\log L$,
equivalent to a cluster luminosity spectrum ${\rm d}N \propto
L^{\alpha} {\rm d} L$, with a spectral index $-2 \la \alpha \la -1.5$
(e.g., Elson \& Fall 1985; Whitmore \& Schweizer 1995; Elmegreen \&
Efremov 1997; Miller et al. 1997; Whitmore et al. 1999; Whitmore et
al. 2002; Bik et al. 2003; de Grijs et al. 2003c; Hunter et al. 2003;
Lee \& Lee 2005; see also Elmegreen 2002). Since the spectral index,
$\alpha$, of the observed CLFs is reminiscent of the slope of the
high-mass regime of the old GC mass spectrum ($\alpha \sim -2$;
McLaughlin 1994), this type of observational evidence has led to the
popular, but thus far mostly speculative, theoretical prediction that
not only a power law, but {\it any} initial CLF (and CMF) will be
rapidly transformed into a Gaussian distribution because of (i)
stellar evolutionary fading of the lowest-luminosity (and therefore
lowest-mass) objects to below the detection limit; and (ii) disruption
of the low-mass clusters due both to interactions with the
gravitational field of the host galaxy, and to internal two-body
relaxation effects leading to enhanced cluster evaporation (e.g.,
Elmegreen \& Efremov 1997; Gnedin \& Ostriker 1997; Ostriker \& Gnedin
1997; Fall \& Zhang 2001; Prieto \& Gnedin 2007).

This approach has two drawbacks, however. Observationally, it has been
pointed out by various authors that for YMCs exhibiting an age range,
one must first correct their CLF to a common age before a realistic
assessment of their {\it initial} CLF (or CMF) can be achieved (e.g.,
Meurer 1995; Fritze--v. Alvensleben 1998, 1999; de Grijs et al. 2001,
2003a,d). Whether the observed power laws of the CLF for YMC systems
are intrinsic to the cluster populations or artefacts caused by the
presence of a cluster age spread -- which might mask a differently
shaped underlying mass distribution -- is therefore a matter of
ongoing debate (see, e.g., Meurer 1995; Fritze--v. Alvensleben 1998,
1999; Carlson et al. 1998; Zhang \& Fall 1999; Vesperini 2000, 2001;
Anders et al. 2007). From a theoretical point of view, while the
preferential removal of the more vulnerable low-mass clusters indeed
results in an initial power-law CMF being turned into an approximately
Gaussian CMF (e.g., Okazaki \& Tosa 1995; Baumgardt 1998; Vesperini
1998; Fall \& Zhang 2001; Prieto \& Gnedin 2007), recovering the {\it
present-day} GCMF after a Hubble time of evolution requires
considerable fine-tuning of the models, which is hardly compatible
with the near-invariance of the GCMF among large galaxies. We review
each of these aspects in Sect. \ref{subsec:IGCMF}.

\subsection{Recovering the Initial GC Mass Function via Modeling}
\label{subsec:IGCMF}

Since individual GCs and GC systems have evolved over a Hubble time in
their galactic environment, the recovery of the initial GCMF is
model dependent. At present, this model dependence is believed to be
through one of two competing hypotheses. The initial GCMF may have
been a featureless power law, with a spectral index $\alpha \sim -2$
(i.e., equivalent to ${\rm d}N \propto m_{\rm cl}^{-2}{\rm d}m_{\rm
cl}$), or a Gaussian distribution similar to that seen today.
Regardless of the actual shape of the initial GCMF, it is worth
bearing in mind that this exercise includes the {\it bound} star
clusters only, as these are the only ones ``recovered'' by the
modeling of the secular dynamical evolution of entire star cluster
systems. In other words, proto-clusters that are disrupted by the time
their member stars reach an age of 10--30 Myr as a result of gas
removal combined with a $\la 30$ per cent SFE are not relevant to the
present discussion.

\subsubsection{Is the Initial GCMF similar to the CLF of YMCs ...}
\label{subsubsec:PL}

In the framework of the initial power-law hypothesis, the Gaussian
distribution characteristic of old GC populations results from
evolutionary effects, predominantly the preferential removal of the
more vulnerable low-mass clusters (Fall \& Zhang 2001). The cluster
mass at the turn-over of the present-day mass function then depends on
the age of the cluster system, and on the cluster disruption
time-scale [see Eqs. (\ref{tdis}) and (\ref{tdis2})], in the sense
that the older the cluster system and/or the shorter the disruption
time-scale, the higher the expected turn-over mass of the cluster
system will be. The evolutionary rate of the GCMF turn-over is thus a
function of the initial spatial distribution of the GCs in their host
galaxy (Parmentier \& Gilmore 2005), of their initial velocity
distribution or, equivalently, of the clusters' orbital distribution
(Murali \& Weinberg 1997b; Baumgardt 1998; Baumgardt \& Makino 2003;
Fall \& Zhang 2001), as well as of the circular (rotation) velocity of
the host galaxy [Baumgardt \& Makino 2003, their eq. 10; see also
Gieles et al. (2006) and Lamers \& Gieles (2007) for the shortening of
the cluster disruption time-scale due to close encounters with giant
molecular clouds]. Thus, the near-invariance of the GCMF in very
different types of galaxies is neither easily understood, nor
straightforwardly reproduced (Vesperini 2001). Additionally, the
predicted cluster mass at the turn-over of model GCMFs is often
significantly lower than observed (Baumgardt 1998; Vesperini
2001). Based on a subsample of host galaxies with effective masses and
radii equal to those determined using observational data for a number
of giant, normal, and dwarf elliptical galaxies, Vesperini's (2001)
fig. 6a shows that the predicted logarithmic GC masses at the
turn-over, at an age of 13 Gyr, are in the range $4.2 \la \langle \log
(m_{\rm cl}[M_{\odot}]) \rangle \la 5$, i.e. significantly lower than
the observed value, $\langle \log (m_{\rm cl}[M_{\odot}]) \rangle \sim
5.3$. Similar results are obtained for the Galactic GC system, with a
predicted $\langle \log (m_{\rm cl}[M_{\odot}])\rangle \sim 4.3$
(Vesperini 1998, his fig. 8). These results were obtained for circular
orbits; we discuss the case of GCs on elliptical orbits below.

Not only is the observed GC mass at the Gaussian peak universal {\it
among} galaxies, it is also universal {\it within} galaxies, i.e. the
turn-over location of the GCMF is constant over a large range of
galactocentric distances (e.g. Harris, Harris \& McLaughlin 1998 and
Kundu et al. 1999 for M87; Kavelaars \& Hanes 1997 and Parmentier \&
Gilmore 2005 for the Galactic halo; Kavelaars \& Hanes 1997 for M31;
Larsen 2006 and Spitler et al. 2006 for the Sombrero galaxy; but see
also Gnedin 1997 for a discussion of the possible role of the
statistical methodology used in the determination of the difference
between the GCLF parameters of the inner and outer clusters).
Equivalently, the shapes of the radial number density profile (the
number of clusters per unit volume in space as a function of
galactocentric distance) and of the radial mass density profile (the
spatial distribution of the cluster system mass around the galactic
centre) of the GC system are similar.

Yet, due to the higher environmental density [see Eq. (\ref{tdis2})]
in the inner region of any GC system, evolutionary processes proceed
at a faster rate there. Therefore, evolutionary models building on the
power-law hypothesis, and assuming an isotropic cluster velocity
distribution, predict a radial gradient of the mean logarithmic
cluster mass (i.e., $\langle \log (m_{\rm cl}[M_\odot]) \rangle$ is
higher in the inner than in the outer regions of a given galaxy). The
resulting gradient appears too large to be consistent with the
observations (Vesperini 2001). Addressing the coherent and
well-defined group of Galactic Old Halo GCs, Parmentier \& Gilmore
(2005) demonstrate that, at an age of 13 Gyr, the theoretically
predicted radial number density profile of the GC system is
significantly shallower than the radial mass density profile, contrary
to observations (see the bottom panels of their figs. 2 and 3). This
behaviour results from the fact that, for a power-law initial GCMF
with a spectral index, $\alpha \sim -2$, low-mass clusters dominate
the total cluster population in terms of number, but not in terms of
mass. Hence, while the radial mass density profile remains
well-preserved throughout their evolution -- in spite of their
preferential disruption -- the radial number density profile becomes
significantly shallower.

Starting from a power-law CMF with a slope similar to that
characterising the CLF of YMC systems in starbursts thus seems to fail
to reproduce fundamental properties of the present-day GCMF. In fact,
the theoretically predicted scatter in the turn-over mass among and
even within galaxies is significantly greater than
observed. Nevertheless, the initial power-law hypothesis cannot be
completely ruled out, although it requires very specific conditions;
careful consideration of the consistency of these conditions with
observational data is required, however. The initial power-law
assumption can account for both the present-day turn-over GC mass and
its near-invariance with galactocentric distance, provided that the
initial GC velocity distribution is characterised by a strong radial
anisotropy that increases as a function of increasing radius, in the
sense that the farther from the galactic centre a given cluster is
located, the higher its orbital eccentricity will be (Fall \& Zhang
2001). The net result of this assumption is that, in these models,
{\it all} GCs have similar perigalactic distances. Because the
mass-loss rate of a GC on an eccentric orbit is significantly more
sensitive to its perigalactic than to its apogalactic distance
(Baumgardt 1998, his fig. 2), the relatively narrow distribution of
pericentres causes a nearly invariant GCMF turn-over over the entire
radial extent of the GC system, regardless of the clusters' loci at
any given point in time. In addition, the small anisotropy radius
($\sim 2$ kpc) of the velocity distribution (or, equivalently, the
mean perigalactic distance) implies that GC evolutionary processes,
and hence the shift of the GCMF towards higher cluster masses, proceed
at a faster rate than in the case of an isotropic velocity
distribution. As a consequence, the cluster mass at the peak of the
GCMF at an age of 13 Gyr matches the observations, both in the inner
and outer galactic regions (Fall \& Zhang 2001).

Vesperini et al. (2003) tested this model against data of the M87 GC
system. Owing to the preferential disruption of GCs on highly
eccentric orbits, the initial amount of radial anisotropy in the
velocity distribution decreases during the evolution of the cluster
population. Yet, the initial radial anisotropy required to reproduce
the near-constancy of the GCMF turn-over mass is so strong that, in
spite of its steady decrease with time, at an age of 13 Gyr, it will
still be significantly stronger than what is inferred from the
observed projected velocity dispersion profile of the M87 GC system
(C\^ot\'e et al. 2001). In other words, the lack of any significant
radial gradient in the mean logarithmic GC mass, and the observed
kinematics of the M87 GC system, cannot be reproduced simultaneously.
The former requires a small initial anisotropy radius ($\la 3$ kpc),
while the latter requires an anisotropy radius greater by at least an
order of magnitude.

In an additional attempt to reconcile the Gaussian mass function of
old GCs with the power-law CLF of YMCs, Vesperini \& Zepf (2003) built
on the observed trend between the mass of Galactic GCs and their
concentration (i.e., the more massive the cluster, the higher its
concentration; van den Bergh 1994; see also Larsen 2006, his fig. 4).
Since this rough correlation is likely of primordial origin (Bellazini
et al. 1996), they investigated how the dissolution of low-mass,
low-concentration clusters induced by early stellar mass loss affects
the temporal evolution of the CMF. Their results suggest that it may
be possible to reproduce, at an age of 13 Gyr, a Gaussian CMF with a
turn-over occurring at about the observed mass, even if starting from
an initial power-law, ${\rm d}N \propto m_{\rm cl}^{-2}{\rm d}m_{\rm
cl}$. Moreover, the dissolution of the low-mass clusters is mostly
caused by their low initial concentration; it depends weakly only on
environmental conditions such as the structure of the host galaxy and
the cluster orbit. Consequently, their model reproduces the lack of
significant radial variations in the mean logarithmic GC mass across
the Galactic GC system, even for an isotropic initial velocity
distribution. Dissolution of low-concentration clusters may therefore
provide the missing link between the power-law CLF observed for YMCs
in violently star-forming environments and the Gaussian CMF
characteristic of old GC populations, although a detailed study of
this effect remains to be done. In addition, accounting for the
near-universality of the GCMF turn-over mass requires the existence,
in other large galaxies containing significant cluster population, of
a similar cluster mass-concentration relationship. This is, as yet,
unknown.

\subsubsection{... or is it similar to the present-day GCMF?}
\label{subsubsec:G}

The large amount of fine-tuning required to reconcile the initial
power-law hypothesis to the observations stimulated the search for an
alternative, more robust, initial GCMF. Building on the {\it N}-body
simulations of Vesperini \& Heggie (1997), Vesperini (1998, 2000)
performed detailed simulations of star cluster systems evolving in
time-independent Milky Way and elliptical galaxy-type gravitational
potentials. He demonstrated the existence of a dynamical equilibrium
Gaussian GCMF. That particular CMF maintains its mean and standard
deviation ($\langle \log(m_{\rm cl}[M_\odot])\rangle \sim 5$ and
$\sigma_{\log m_{\rm cl}} \sim 0.6$ dex) during the entire evolution
through a subtle balance between disruption of clusters on the one
hand and evolution of the masses of the surviving clusters on the
other, even though a significant fraction of the clusters is
destroyed. In addition, owing to the combined effects of tidal
disruption and dynamical friction (which preferentially affect
low-mass and high-mass clusters, respectively), {\it any} Gaussian
ICMF eventually evolves into that equilibrium CMF (Vesperini 1998, his
fig. 13). The speed at which this evolution proceeds is determined by
the initial deviation of the system from the equilibrium CMF.
 
The shape of the equilibrium CMF is very close to that of the GCMF
observed in the outer regions of elliptical galaxies, where the
initial conditions are likely retained because of the low efficiency
of cluster disruption processes expected at large galactocentric
distances (Vesperini 2000; see also McLaughlin, Harris \& Hanes 1994).
Therefore, these GCMFs may be significantly better proxies to the
initial CMF than the featureless power-law (Murali \& Weinberg
1997a,b). As a dynamical equilibrium CMF, its shape is practically
independent of both the age of the cluster system and the cluster
disruption time-scale. Its temporal evolution therefore differs
substantially from that of the initial power-law CMF, as the turn-over
mass of the latter depends heavily on initial and environmental
properties. In contrast, the equilibrium GCMF maintains its initial
shape during its evolution, regardless of either the underlying
initial cluster position-velocity distribution, or the rotation
velocity of the host galaxy. As a result, the near universality of the
GCMF turn-over mass and the small radial dependences observed in GC
systems with sufficient radial sampling (see Sect.
\ref{subsubsec:PL}, and references therein) are retrieved naturally.
Only the fraction of surviving clusters and the ratio of the final to
initial total mass in clusters vary among and within galaxies, as a
result of cluster disruption time-scale variations.

Generally speaking, the evolution of other bell-shaped GCMFs with
similar turn-over masses (e.g., a Student $t$-distribution; Secker
1992) is not markedly different from that of the equilibrium Gaussian
GCMF (Vesperini 2002; see also Parmentier \& Gilmore 2007). The
bell-shaped initial GCMF generated by the gas-removal model of
Parmentier \& Gilmore (2007) shows, once corrected for stellar
evolutionary mass losses, a mean logarithmic cluster mass at about a
factor of 2 higher than that of Vesperini's (1998, 2000) equilibrium
Gaussian CMF (i.e., $\langle \log (m_{\rm cl}[M_\odot]) \rangle \sim
5.3$), as well as a more prominent low-mass wing than expected for a
Gaussian CMF. In spite of these differences, once evolved to an age of
13 Gyr using the results of the {\it N}-body simulations performed by
Baumgardt \& Makino (2003), this initial GCMF also appears to be an
almost perfect dynamical equilibrium GCMF (Parmentier \& Gilmore 2007,
their fig. 11).

Although considerable work has been done on this topic, the shape of
the initial GCMF still remains a model-dependent and contentious
issue. We note that, while the initial power-law cannot be firmly
ruled out (Vesperini \& Zepf 2003), a Gaussian initial GCMF similar to
the universal present-day GCMF represents a safer solution by virtue
of its robustness to model inputs.

\subsubsection{The 1-Gyr old fossil starburst site M82 B}
\label{subsubsec:M82B}

Based on deep {\sl HST} optical and near-infrared imaging, de Grijs et
al. (2003a,d) reported the first discovery of an approximately
Gaussian CLF (and CMF) for the roughly coeval, $\sim 1$ Gyr-old star
clusters in the fossil starburst region ``B'' of M82. With a turn-over
mass at $\log (m_{\rm cl} [M_\odot]) = 5.1 \pm 0.1$ (which occurs some
2 mag brighter than the 100 per cent observational completeness; this
corresponds to $\log (m_{\rm cl} [M_\odot]) \simeq 4.4$), this CMF
very closely matches the universal GCMF of old GC systems (de Grijs et
al. 2003d, their fig. 1c). The fact that they considered an
approximately coeval subset of the M82 B cluster population, combined
with the use of the 100 per cent completeness limit as their base line
ensures the robustness of the CMF peak detection (see de Grijs et
al. 2005b for a detailed discussion). This provided the first deep CLF
(CMF) for a star cluster population at intermediate age, which thus
serves as an important benchmark for theories of the evolution of star
cluster systems. In this respect, the M82 B cluster population
represents an ideal sample to test these evolutionary scenarios since,
for such a roughly coeval intermediate-age population in a spatially
confined region, (i) the observational selection effects are very well
understood (de Grijs et al. 2003a,d), and (ii) the dynamical cluster
disruption effects are very similar for the entire cluster population.

{\it If} the ICMF is a power-law with a spectral index, $\alpha \sim
-2$, a cluster mass at a 1-Gyr old turn-over similar to that observed
for 13-Gyr old GC populations implies that, in the M82 B region,
cluster evolution has {\it necessarily} proceeded at a much faster
rate than in, e.g., the Galactic halo. This is required in order to
deplete the star cluster system at an accelerated rate, and evolve the
presumed initial power law to the observed distribution on a
time-scale of $\sim 1$ Gyr. In fact, using the Boutloukos \& Lamers
(2003) framework, de Grijs et al. (2003a) deduce that in order to
produce a Gaussian present-day CMF in M82 B as is observed, the
time-scale on which a typical $\sim 10^4 M_\odot$ cluster is expected
to disrupt must be $t_4^{\rm dis} \simeq 30$ Myr. This is the shortest
disruption time-scale known in any (disk region of a) galaxy. In
addition, de Grijs et al. (2005b) noticed that such a short a
disruption time-scale implies a correspondingly low cluster survival
rate, implying that the initial number of clusters, $\sim 80,000$ with
an average mass of order $m_{\rm cl} \sim 10^4 M_\odot$, would be
unphysically high for the relatively small ($\la 1$ kpc$^3$) M82 B
region. Observations of the present-day M82 B CMF are thus
inconsistent with a scenario in which the 1 Gyr-old cluster population
originated from an initial power-law mass distribution.
\footnote{We caution that the M82 B cluster mass estimates derived by
de Grijs et al. (2003a) are based on a Salpeter (1955)-type stellar
IMF. Recent determinations of realistic stellar IMFs deviate
significantly from that representation at low masses, in the sense
that they appear to be significantly flatter than the Salpeter slope.
The implication of this is, therefore, that we may have {\it
overestimated} the individual cluster masses. A more modern IMF, such
as that of Kroupa, Tout \& Gilmore (1993) implies overestimated
cluster masses by factors of 1.7 to 3.5 for an IMF containing stellar
masses in the range $0.1 \le m_\ast/M_\odot \le 100$, the exact value
depending on the slope adopted for the lowest stellar mass range,
$0.08 < m_\ast/M_\odot \le 0.5$. This corresponds to a correction of
$-0.23$ to $-0.54$ to the peak of the CMF, so that a more realistic
estimate for the peak of the M82 B CMF would be $\langle \log(m_{\rm
cl} [M_\odot]) \rangle = 4.7 \pm 0.2$.  However, once again, the
observed cluster population would be the remains of an unphysically
high initial census of $\sim 20,000$ clusters if starting from an
initial power-law mass distribution.} Alternatively, considering the
close coincidence in shape of the observed CMF to that of Vesperini's
(1998) (quasi-)equilibrium CMF, they demonstrate that the M82 B ICMF
may have been a Gaussian similar to that of old GCs, irrespective of
the exact cluster disruption time-scale. The analysis of the M82 B
cluster sample, and the conclusions drawn from these observations,
have proven to be contentious. With the recent release of the M82 {\sl
HST}/ACS Treasury data set of imaging observations in the F435W, F555W
and F814W broad-band filters as well as in the F658N H$\alpha$ filter,
supplemented by targeted ACS/High-Resolution Channel (HRC)
observations (GO-10853; PI L. J. Smith) of M82 B in particular in the
F330W (equivalent to the Johnson $U$-band filter) should shed further
light on the robustness of the CMF peak in this region. Initial
results (L. J. Smith, priv. comm.) suggest that the peak of the age
distribution inferred from the $U$-band selected YMCs in M82 B is
shifted to systematically somewhat lower ages than those obtained by
de Grijs et al. (2001, 2003a) without access to these short
wavelengths.

\subsubsection{The merger remnant NGC 1316}
\label{subsubsec:NGC1316}

Recently, Goudfrooij et al. (2004) added a second important data point
to constrain cluster evolution theories, based on the $\sim 3$ Gyr-old
metal-rich ($Z \sim Z_\odot$) cluster population in NGC 1316, a merger
remnant, based on {\sl HST}/ACS observations. The CLF of the
intermediate-age population of massive star clusters in NGC 1316
appears to be a power-law with a spectral index, $\alpha \simeq
-1.75$, down to the observational 50 per cent completeness limit. They
divide their cluster sample into two equal-sized subsamples sorted by
projected galactocentric distance. Whereas they detect a clear
turn-over at $M_V \sim -6.2$ mag in the ``inner'' cluster population
($R_{\rm proj} \le 9.4$ kpc), the CLF of the ``outer'' population
continues to rise all the way down to the detection limit. If the ICMF
in this galaxy were a power-law similar to that of the ``outer''
population, this illustrates the greater efficiency of evolutionary
processes at closer distance from the galactic centre, as well as the
absence of any strong radially dependent velocity distribution
anisotropy. To compare the observed CLFs with the CMFs predicted by
Fall \& Zhang (2001), Goudfrooij et al. (2004) convert the latter into
CLFs on the basis of Maraston et al.'s (2001) stellar population
synthesis models. They note that the turn-over magnitude of the
``inner'' CLF matches Fall \& Zhang's (2001) predictions rather well
at an age of 3 Gyr (see their fig. 3f, although the observed width of
the CLF is significantly narrower than predicted). They deduce that
the ``inner'' CLF (and CMF) will eventually evolve into that
characteristic of old GCs by the time the NGC 1316 ``inner'' clusters
reach an age of 13 Gyr.  Clearly, this conclusion does not apply to
the ``outer'' clusters. The power-law CLF of the latter subsample
illustrates that these clusters have thus far experienced little, if
any, preferential depletion of low-mass clusters, and hence dynamical
evolution (at least above the observational completeness limit).

However, de Grijs et al. (2005b) note that based on the published CLFs
and the discussion in Goudfrooij et al. (2004), and assuming a
Salpeter-like IMF with masses $m_\ast \ge 0.1$ M$_\odot$, the {\sc
galev} simple stellar population models (Schulz et al. 2002; Anders \&
Fritze--v. Alvensleben 2003) indicate a mean cluster mass of
$\log(m_{\rm cl}/M_\odot) \sim 4.0$, with a FWHM of $\sim 0.9$
dex. These are significantly smaller masses (and a smaller width) than
expected for GC progenitors. At an age of 13 Gyr, the cluster mass at
the ``outer'' CMF turn-over will thus be markedly lower than the
universal turn-over GC mass of $2 \times 10^5 M_\odot$. The difference
in shape between the CLFs of the ``inner'' and ``outer'' samples also
demonstrates that the solar-abundance intermediate-age massive
clusters of NGC 1316 fail to reproduce the near-invariance of the mean
logarithmic cluster mass with galactocentric distance that is observed
for the ubiquitous old GC systems. Therefore, while at least a
fraction of the observed clusters will likely survive until an age of
13 Gyr, particularly in the outer regions, they do not represent
proto-GCs in the sense that their CMF does {\it not} show the
properties characteristic of the universal GCMF.

\subsection{The CMF of young massive clusters and the (many) associated 
caveats}
\label{sec:young_SC}

\subsubsection{Caveats beyond the Local Group}
\label{sec:bLG}

Observational evidence of YMC systems in many star cluster-forming
interacting and starburst galaxies appears to indicate that they are
well represented by power-law cluster {\it luminosity} functions with
a spectral index in the range from $-1.8$ to $-2$ (see Sect.
\ref{subsec:CLF_facts} and references therein; de Grijs et al. 2003c
for a comprehensive review; but see also Anders et al. 2007). Yet,
rapid changes to the properties of young stellar populations, combined
with a possible age range within a given cluster system (possibly on
the order of its median age) may conspire to give rise to a
power-law-like CLF, even if the true underlying cluster {\it mass}
function is not a power-law (Meurer 1995; Fritze--v. Alvensleben 1998,
1999; de Grijs et al. 2001, 2003a,d; Hunter et al. 2003). In addition,
because of the fading with time of the cluster luminosity, high-mass
clusters can be observed over a wide range of ages, while low-mass
clusters are detectable at young ages only, for a flux limited sample
selection. This results in the underrepresentation of the latter in
the age-integrated CMF. Therefore, it is obviously very important to
age-date the individual clusters, before interpreting the cluster
luminosities in terms of the corresponding mass distribution.

Beyond addressing the issue of the shape of the CMF in violently
star-forming environments, and because the old GCMF itself appears to
be universal, it is worth considering whether the ICMF of present-day
forming massive star clusters is near-invariant as well. However, the
CMF of young massive star clusters (which we take as a good proxy to
the ICMF in the absence of any significant secular dynamical
evolution) has been probed to sufficient depth in only a limited
number of galaxies.

In most cases, the reported CMF is consistent with a power-law of
which the spectral index has the canonical $\alpha \sim -2$ value.
The depth of the cluster mass range probed by the observations varies
among galaxies, however. Bik et al. (2003) infer a power-law CMF down
to $10^3 M_\odot$ for the cluster population in the inner spiral arms
of the interacting galaxy M51 (see also the discussion in
Sect. \ref{m51.sec}). Similar results are obtained by Zhang \& Fall
(1999) for the YMCs of the Antennae system for cluster age ranges $2.5
< t < 6.3$ Myr and $25 < t < 160$ Myr, down to the observational
completeness limits, at $\simeq 8 \times 10^3 M_\odot$ and $\simeq 25
\times 10^3 M_\odot$, respectively. In contrast to Zhang \& Fall
(1999), however, Fritze--v. Alvensleben (1998, 1999) favours a Gaussian
mass distribution at young age, similar to the present-day GCMF.
Fritze--v. Alvensleben (2004) notes that both approaches have their
drawbacks. Zhang \& Fall (1999) exclude a significant number of
clusters from the ambiguous age range in the reddening-free $Q_1-Q_2$
index diagram they use, while Fritze--v. Alvensleben (1998, 1999)
assumes a uniform average reddening in the {\sl HST} / WF/PC1 {\it
UVI} data. A recent re-analysis of {\sl HST}/WFPC2 data of the
Antennae galaxies (Anders et al. 2007) may help clarify this complex
situation (see below). If one accepts the hypothesis that the ICMF of
old GCs is a Gaussian similar to what is observed today, these results
suggest that the ICMF of present-day massive star clusters differs
from that of old GCs. Similar power-law CMFs have been reported by de
Grijs et al.  (2003c) for the cluster populations in NGC 3310 and NGC
6745. Yet, in those both cases, the cluster mass range coverage
coincides only with the high-mass regime of the Gaussian GCMF
(i.e. $>10^5 M_\odot$ and $> 4 \times 10^5 M_\odot$, respectively), so
that one cannot distinguish between power-law and Gaussian CMFs for
these cluster samples. In contrast to the featureless power-laws
uncovered for M51 and the Antennae system, Cresci, Vanzi \& Sauvage
(2005) report the existence of substructures in the CMF of the YMC
population in the central region of NGC 5253, an irregular dwarf
galaxy in the Centaurus Group, observed with {\sl HST}/ WFPC2. The
derived cluster ages span a range between 3 and 20 Myr. They note that
the ``young'' and ``old'' clusters\footnote{They distinguish between
young and old clusters on the basis of their presence or absence in
H$\alpha$.} of their overall sample show markedly different
CMFs. While the ``young'' population (with mean age $\simeq 8$ Myr)
shows a power-law CMF with a spectral index, $\alpha \simeq -1.6$, the
``old'' clusters (to which they assign a reference age of $\sim 20$
Myr) display a turn-over at about $5 \times 10^4 M_\odot$, i.e., at a
mass some 2.5 times more massive than their 50 per cent completeness
limit. Note that according to Fall \& Zhang (2001), such a high
cluster mass at the CMF turn-over is predicted for a much older age,
of 2 Gyr, if one starts from an initial power-law CMF. Considering the
very young age of the NGC 5253 clusters, the observed turn-over mass
is likely not an imprint of ongoing dynamical evolution, but instead
probably a trace of their formation process. Beyond the turn-over
mass, the CMF is consistent with a power-law of spectral index,
$\alpha \simeq -1.8$, i.e. in agreement with what is inferred for
other YMC systems. As for the ``young'' clusters, their shallower
spectral index, $\alpha \simeq -1.6$, is more reminiscent of what is
observed for clumps and star-forming cores in giant molecular clouds
(see, e.g., Lada \& Lada 2003).

In summary, compact YMCs show power-law CMFs with spectral index
$\alpha \la -2$. Whether this behaviour characterises the entire
cluster mass range down to, say, $100 M_\odot$, as observed for the
less massive open clusters in the Galactic disc (Lada \& Lada 2003;
and references therein; see also the catalogue by Kharchenko et
al. 2005) remains to be seen, however. NGC 5253 (Cresci et al. 2005)
and M82 B (de Grijs et al. 2005b) seem to constitute cases for which
this is actually not the case. Deeper observations for a larger sample
of galaxies are, as always, required.

Finally, prior to proceeding any further, we note that results
reported for the CLF itself differ sometimes significantly among
studies. It is worth bearing in mind that retrieving the {\it
intrinsic} CLF requires avoiding bright-star contamination and
accounting for completeness effects properly. This may prove
challenging at the faint end of the luminosity distribution. Existing
CLF substructures at low luminosities may therefore remain undetected
or lead to discrepant results. For instance, while Whitmore et
al. (1999) infer from their {\sl HST}/WFPC2 imaging data a power-law
CLF with a spectral index $\alpha \simeq -2.1$ for the YMCs in the
Antennae system, Anders et al. (2007) report the first statistically
robust detection of a turn-over in its CLF at $M_V \simeq -8.5$
mag. The origin of that discrepancy likely resides in differences in
the data reduction and the statistical analysis aimed at rejecting
bright-star contamination and disentangling completeness effects from
intrinsic CLF substructures.

\subsubsection{The Magellanic Clouds: test case on our door step}
\label{subsubsec:MC}

The preceding discussion has illustrated that it is not necessarily
straightforward to derive the mass distributions of YMC systems beyond
the Local Group. In contrast, the Magellanic Clouds are close enough
for a detailed survey of even faint clusters. To derive the CMF of
mass-limited LMC cluster subsamples, which are potentially more
physically informative than magnitude-limited subsamples, de Grijs \&
Anders (2006) re-analyse the {\sl UBVR} broad-band SEDs of Hunter et
al. (2003). Building on the framework established by Boutloukos \&
Lamers (2003), they derive that the time-scale on which a $10^4
M_\odot$ cluster is expected to disrupt is $\log(t_4^{\rm dis}{\rm
yr}^{-1})=9.9\pm0.1$ [see also Eq. (\ref{tdis})], which is in
agreement with previous determinations of the characteristic cluster
disruption time-scale. Such a long cluster disruption time-scale
results from the low-density environment of the Magellanic Clouds. It
guarantees that the observed cluster mass distributions have not yet
been altered significantly by secular dynamical evolution,
i.e. clusters already affected by ongoing disruption have faded to
below the completeness limit (see de Grijs \& Anders 2006, their
fig. 8). As a result, the observed mass distributions are the {\it
initial} distributions. Considering clusters older than 60 Myr and
more massive than $10^3 M_\odot$, the spectral index of their CMF is
$\alpha \simeq -2$, i.e. fully consistent with what is generally
inferred for YMC systems. This slope prevails over the age range 100
Myr--6 Gyr (see also de Grijs et al. 2007). Similar results were
obtained by Hunter et al. (2003). However, the spectral index of the
clusters younger than 60 Myr appears significantly shallower, with
$\alpha \simeq -1.7$ for the same mass range and $\alpha \simeq -1$
for clusters less massive than $10^3 M_\odot$. As discussed in Sect.
\ref{infant.sec}, the origin of these substructures in the young
age/low-mass regime is likely to be found in a mass-dependent infant
mortality rate, at least for the lowest-mass cluster range. It is
important to bear in mind, however that at such a young age, the
population of clusters necessarily consists of a mix of bound and
unbound clusters. Most of these ($\sim 60$--90 per cent) will disperse
by the time their member stars reach an age of a few $\times 10^7$ yr
(see Sect. \ref{infant.sec}, and references therein).

\section{Summary and Outlook}
\label{summary.sec}

The formation of GCs, which was once thought to be limited to the
earliest phases of galaxy formation, appears to be continuing at the
present time in starburst, interacting and merging galaxies in the
form of star clusters with masses and compactnesses typical of GCs.
Whether these YMCs will evolve to become old GCs by the time they
reach an age of 13 Gyr depends to a very large extent on their
environment, however. For a host galaxy with a smooth logarithmic
gravitational potential, the ambient density seems to be the key
parameter driving the rate of cluster evolution. This is accelerated
in the presence of substructure in the host galaxy, such as that
commonly provided by bulge, spiral arm and giant molecular cloud
components.

The LMC and the disk of M51 represent strikingly different cases of
the cluster ability to survive to old age. The M51 disk is
characterised by a cluster disruption time-scale $\rm t_{\rm dis}^4
\sim 100$ Myr (Gieles et al. 2005), which implies that even a $10^6
M_\odot$ cluster will not survive much longer than $\sim 2$ Gyr. In
contrast, the more gentle environment of the LMC is conducive to a
long cluster disruption time-scale, $\rm t_{\rm dis}^4 \sim 8$ Gyr (de
Grijs \& Anders 2006), so that the same $10^6 M_\odot$ cluster will
survive largely unaffected, except for $\sim 30$ per cent stellar
evolutionary mass loss, over the next Hubble time. Therefore, although
survival rates vary significantly as a function of the star cluster
system host properties, at least a fraction of the observed YMCs that
have recently been formed in starburst environments will likely reach
an age typical of that of old GCs.

In this respect, a remaining contentious issue is whether the observed
CMF of YMCs will eventually evolve into that of the ubiquitous old
GCs. The GCMF is a Gaussian with a mean $\langle \log(m_{\rm
cl}[M_\odot]) \rangle \sim 5.2-5.3$ and a standard deviation of
$\sigma_{\log m_{\rm cl}} \simeq 0.5-0.6$ dex. It seems to be almost
universal, both {\it among} and {\it within} galaxies. On the other
hand, many CMFs of YMCs appear to be featureless power laws with a
spectral index $\alpha \sim -2$ down to a few $\times 10^3 M_\odot$
(e.g., de Grijs \& Anders 2006; Hunter 2003, for the LMC). Some
cluster systems exhibit differently shaped ICMFs, however (M82 B, de
Grijs et al. 2005b; NGC 1316, Goudfrooij et al. 2004; NGC 5253, Cresci
et al. 2006).

Evolving such an initial power-law CMF into the near-invariant
Gaussian GCMF regardless both of the host galaxy properties and of the
details of the cluster loci turns out to be most challenging and
requires significant fine-tuning of the models, which is not
necessarily compatible with the available observational constraints
(see, e.g., Fall \& Zhang 2001 vs. Vesperini et al. 2003; see also
Vesperini \& Zepf 2003). Nonetheless, it is worth bearing in mind that
present-day GC evolutionary models describe the host galaxy as a {\it
static} gravitational potential. This is most likely an
oversimplification when addressing the evolution of star clusters that
formed when galaxies were still in the process of being assembled.

Since the present-day CMF appears to be close to the dynamical
equilibrium shape (Vesperini 1998), the initial GCMF may also have
been a Gaussian similar to that seen today. This GCMF preserves its
shape during its evolution, irrespective of the cluster disruption
time-scale, and hence of the ambient density. Consequently, the main
properties of the GCMF, i.e. the near-invariance among and within
galaxies, are retrieved naturally.

Therefore, even though at least a fraction of the observed YMCs will
likely evolve into old GC counterparts by the time they reach an age
of 13 Gyr (albeit of higher metallicity), GCs and YMCs may differ with
respect to the shape of their ICMF. Parmentier \& Gilmore (2007)
recently explored how the gas-removal phase affects the mapping of the
mass function of the gaseous cluster progenitors to the ICMF. As they
assume that the SFE is mass independent, which is probably a
reasonable assumption for cluster-forming clouds more massive than
$10^4 M_\odot$ (Bastian et al. 2005; Fall et al. 2005; de Grijs et
al. 2007), they show that a featureless power-law cloud mass function
evolves into a featureless power-law of the same slope (because of the
assumed independence of the SFE on cloud mass). However, as soon as
the power-law mass function of the cluster parent molecular clouds
shows a lower mass limit, a bell-shaped ICMF is generated as a result
of gas removal, with the turn-over located at a cluster mass on the
order of the lower mass limit of the clouds. Specifically, they
demonstrate that, if the proto-GC cloud mass distribution is
characterised by a mass scale of $10^6 M_\odot$, then {\it any}
proto-GC cloud mass function (e.g., a Gaussian with a mean of $\sim
10^6 M_\odot$ and $\sigma_{\log m_{\rm cl}} \la 0.4$ dex, or a power
law truncated at $\sim 6 \times 10^5 M_\odot$) evolves into an initial
GCMF with the appropriate turn-over mass (although the detailed shape
of the initial GCMF actually depends on the shape of the proto-GC
cloud mass function). Therefore, with the mass scale of the cluster
gaseous progenitors as its key parameter, Parmentier \& Gilmore's
(2007) model can account for both the Gaussian initial GCMF and the
observed power-law mass spectrum of YMCs formed in starbursts and
mergers.

The suggestion that the ICMFs of old GCs and of YMCs may be different
is also put forward by Hunter et al. (2003). As they fit cluster
population models to the LMC cluster mass distribution integrated over
age and to their age distribution integrated over mass, they note that
the power-law ICMF with a spectral index $\alpha \ga -2$ which they
determine for the young and intermediate-age clusters fails to
reproduce the distributions of the oldest clusters, i.e. those similar
to Galactic halo GCs in terms of their ages and masses. This suggests
a differently shaped ICMF for the old massive LMC clusters, although
the limited number of clusters necessarily hampers the significance of
this result.

Regardless of the remaining uncertainties with respect to the shape of
the initial GCMF, it now appears that, once evolved to an age of 13
Gyr, YMCs will not exhibit a similar universal CMF as the ubiquitous
old GCs (see also the results of Goudfrooij et al. 2004 for the NGC
1316 CLF and CMF as a function of galactocentric radius). This is due
to the vastly different cluster disruption time-scales characterising
the various environments in which they are found (e.g., Lamers et
al. 2005, their fig. 4), and hence to the different rates at which the
evolved CMF turn-over mass evolves towards larger cluster masses as a
result of the preferential removal of low-mass clusters.

In order to settle the issues of cluster evolution and ICMF shape more
conclusively, major improvements are required in the near future, both
observationally and theoretically. Observations reaching low-mass
clusters, and with sufficiently accurate photometry, in order to
derive reliable cluster ages, are required to follow the temporal
evolution of the CMF. Of specific interest is the presence and mass of
a turn-over, as this will provide an estimate of the cluster
disruption rate. It is also worth following in detail the first $\sim
50$ Myr of cluster evolution, in order to better probe the process of
infant mortality and infer its possible cluster mass dependence for
the lowest-mass range. From a modeling point of view, a better
treatment of the initially loosely bound clusters (i.e., the
low-concentration clusters) is required, since these may account for
the missing link between the Gaussian GCMF and the power laws seen for
YMC systems (Vesperini \& Zepf 2003). In addition, the inclusion of a
time-dependent host galaxy gravitational potential will enable us to
better follow the early evolution of both old GCs and YMCs formed in
interacting and merging galaxies.

\begin{acknowledgements}
We would like to thank Prof. J. X. Wang for the invitation to write
this review. We thank Simon Goodwin and Nate Bastian for insightful
discussions during the preparation of this manuscript, and for
providing us with the data points used for Fig.  \ref{diagnostic.fig},
and Paul Crowther for useful insights. We thank Remco Scheepmaker for
making the M51 cluster age and mass determinations available to us in
electronic format. We also acknowledge research support from and
hospitality at the International Space Science Institute in Berne
(Switzerland), as part of an International Team programme. GP
acknowledges support from the Belgian Science Policy Office in the
form of a Return Grant. This research has made use of NASA's
Astrophysics Data System Abstract Service. We acknowledge reference to
the literature survey by Sarah Moll, done as part of the requirements
to obtain a Ph.D. degree at the University of Sheffield.
\end{acknowledgements}

\label{lastpage}

\end{document}